\documentclass[preprint2,twocolappendix]{aastex6}
\usepackage{amsmath}

\begin{document}

\title{ Uniform Silicon Isotope Ratios Across the Milky Way Galaxy}

\author{Nathaniel N. Monson\altaffilmark{1}, Mark R. Morris\altaffilmark{2}, and Edward D. Young\altaffilmark{1}}
\affil{ $^1$Department of Earth, Planetary, and Space Sciences, UCLA, $^2$Department of Physics \& Astronomy, UCLA}

%%\altaffiltext{1}{Department of Earth, Planetary, and Space Sciences, UCLA}
%%\altaffiltext{2}{Department of Physics and Astronomy, UCLA}

\begin{abstract}
We report the relative abundances of the three stable isotopes of silicon, $^{28}$Si, $^{29}$Si and $^{30}$Si, across the Galaxy using the $v = 0, J = 1 \to 0$ transition of silicon monoxide. The chosen sources represent a range in Galactocentric radii ($R_{\rm GC}$) from 0 to 9.8 kpc. The high spectral resolution and sensitivity afforded by the GBT permit isotope ratios to be corrected for optical depths.  The optical-depth-corrected data indicate that the secondary-to-primary silicon isotope ratios $^{29}{\rm Si}/^{28}{\rm Si}$ and $^{30}{\rm Si}/^{28}{\rm Si}$ vary much less than predicted on the basis of other stable isotope ratio gradients across the Galaxy.  Indeed, there is no detectable variation in Si isotope ratios with $R_{\rm GC}$.  This lack of an isotope ratio gradient stands in stark contrast to the monotonically decreasing trend with $R_{\rm GC}$ exhibited by published secondary-to-primary oxygen isotope ratios.  These results, when considered in the context of the expectations for chemical evolution, suggest that the reported oxygen isotope ratio trends, and perhaps that for carbon as well, require further investigation. The methods developed in this study for SiO isotopologue ratio measurements  are equally applicable to Galactic oxygen, carbon and nitrogen isotope ratio measurements, and should prove useful for future observations of these isotope systems.
\end{abstract}


%\cite{
%\citep{

\section{Introduction}

The utility of interstellar isotope abundance ratios as diagnostic tools for probing metallicity variations across the Galaxy was realized well over thirty years ago \citep[e.g.]{Linke:1977,Frerking:1980,Penzias:1981,Penzias:1981a,Wilson:1981,Wolff:1980}. In conjunction with models for Galactic chemical evolution (GCE), the distribution of stable isotopes  with distance from the Galactic center provides a quantitative probe into stellar nucleosynthesis\citep{Henkel:2014,Prantzos:1996,Timmes:1995}, galaxy formation and evolution\citep{Kobayashi:2006,Martin:2009,Martin:2010,Prantzos:1996,Spite:2006} and levels of heterogeneity in the interstellar medium (ISM) \citep{Lugaro:2003,Nittler:2005,Young:2011}. For these purposes, Galactocentric radius ($R_{\mathrm{GC}}$, i.e. distance from the Galactic center) serves as a proxy for time  because stellar processing of material increases with both decreasing $R_{\mathrm{GC}}$ and time.

 Galactic chemical evolution of light stable isotopes leads to shifts in isotope ratios over time in what should be broadly predictable ways.  The shifts are especially pronounced for ratios of secondary nuclides to primary nuclides, and the details of the process are clearer where two or more such ratios are available.  When studied as functions of $R_{\mathrm{GC}}$, isotopic abundance ratios delineate the extent of stellar processing within the Galaxy, and serve as signposts for chemical variations with time \citep{Clayton:1984,Clayton:1986,Timmes:1995,Prantzos:1996,Prantzos:2008,Kobayashi:2011}.  

The ratios of secondary to primary silicon isotopes in the solar system are surprisingly low compared with older presolar SiC grains found in meteorites. This aberration has been used as possible evidence for extraordinary enrichment of the primary isotope $^{28}$Si by supernovae in the region in which the Sun formed \citep{Alexander:1999, Young:2011}. In order to verify or contravene the idea that the birth environment of the solar system was atypical of the Galaxy 4.6 Gyr before present, one needs an understanding of how the relevant stable isotope ratios have evolved with time and place in the Galaxy (i.e., over the last 4.6 Gyrs).  We need to understand whether  our solar system formed from typical material and by typical processes, or, whether it formed in some atypical environment and/or by unusual processes. In other words, \emph{are we normal} in the context of the isotopic evolution of our local Galactic environs?  
The solar system is expected to be representative of the  interstellar medium (ISM) at $R_{\mathrm{GC}} \approx$ 8 kpc, 4.6 Gyr before present, in the absence of some extraordinary local enrichment processes during its formation.  GCE  over this time interval must be accounted for before drawing comparisons between the solar system and the present-day ISM in a meaningful way. Studies of isotope ratios vs. $R_{\mathrm{GC}}$ therefore provide the context for interpreting the significance of solar-system stable isotope ratios. If our solar system fits with the general picture of secular variations in stable isotope ratios in the Galaxy, then it would suggest that the answer to this question is at least in part in the affirmative. Conversely, if our solar system exhibits significant departures from the averages expected from an analysis of the distribution and evolution of isotopes in the Galaxy, then we will be impelled to search for extraordinary circumstances to explain these departures in isotopic abundances (enrichment by nearby supernovae is the most obvious example). Isotopes of silicon are thought to be an example of the latter case but a firm Galactic reference frame for interpretation of the solar data is not in place. 

Studies of isotope ratios vs.\ Galactocentric radius therefore help place the solar system in a Galactic perspective, and provide the context for interpreting the significance of solar system stable isotope ratios.  This is the objective of the present study.  The first step is to establish the baseline isotopic characteristics of the Galaxy. This in turn  involves defining the mean distributions of isotope ratios as functions of $R_{\mathrm{GC}}$, and establishing the magnitude of dispersion about this trend.  

Rare stable isotopes often comprise only a percent or less of the total abundance of the element of interest. As a result, signal-to-noise ratios (SNRs) for emission lines from rare isotopologues are typically poor and contribute significantly to the error budgets. Measurements of the abundance ratios of the three stable isotopes of silicon by \cite{Wolff:1980} and soon after by \cite{Penzias:1981}, using the $v = 0, J = 2 \to 1$ and $J = 3 \to 2$ lines of SiO, were hampered by low signal-to-noise.  However, modern cryogenic HEMT amplifiers and SIS mixers provide such exceptionally low noise that sensitivities have been increased in excess of an order of magnitude since those early studies. The data reported by \cite{Penzias:1981} and \cite{Wolff:1980} have statistical errors as high as $40\%$. The measurements of  [$^{29}$Si]/[$^{28}$Si] and [$^{30}$Si]/[$^{28}$Si] ratios based on the $v = 0, J = 1 \to 0$ transitions of SiO reported herein have $1\sigma$ statistical errors one tenth of that value. In part for this reason, stable isotope abundance ratios as tracers for variations in the degree of astration across the Galaxy should see a resurgence  \citep[e.g., ][]{Adande:2012,Henkel:2014}.

\section{Previous Work}
\subsection{Stellar Metallicity}

\begin{figure}[t]
  \begin{center}
    \includegraphics[width=0.48\textwidth]{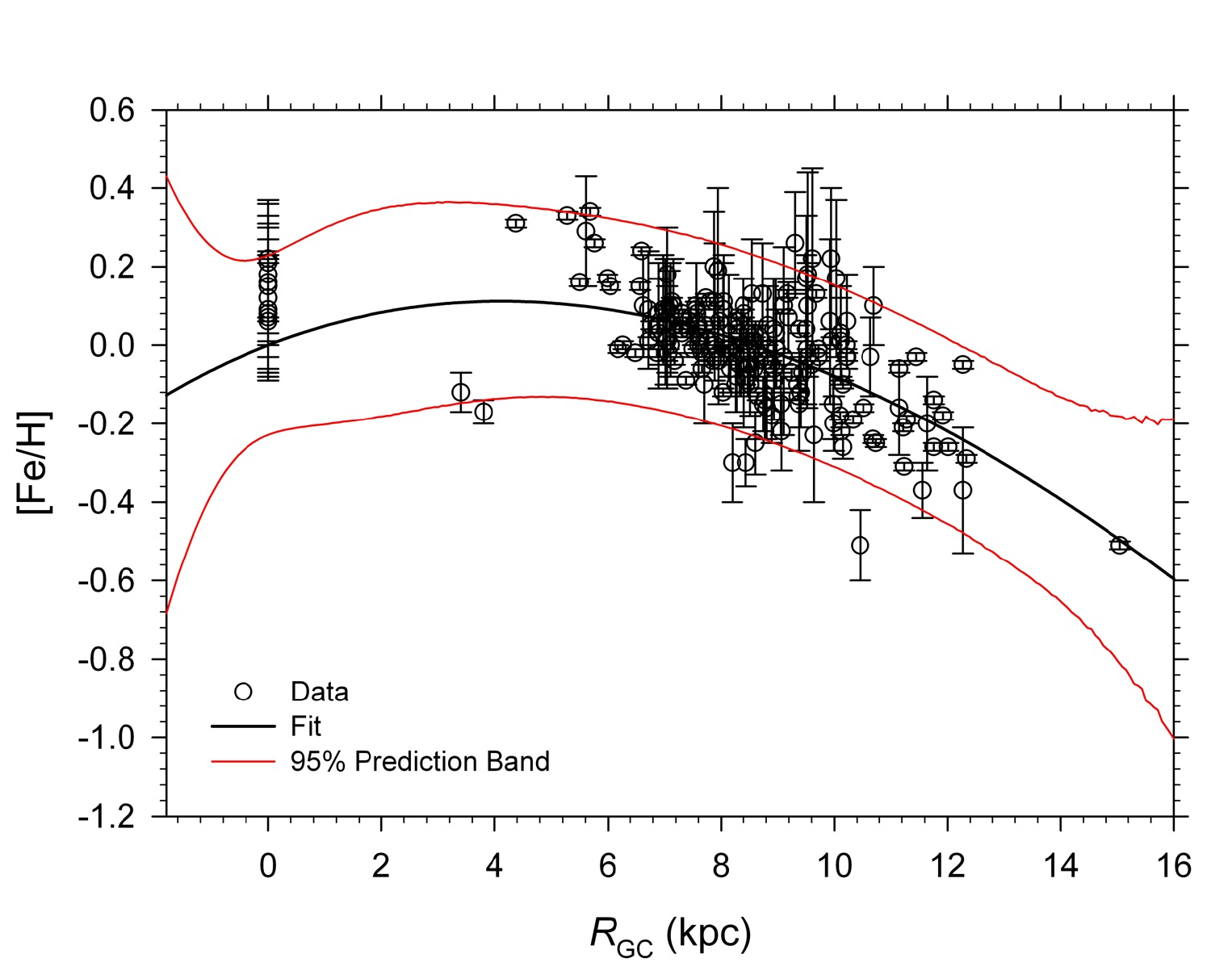}
  \end{center}
  \vspace{-10pt}
  \caption{Stellar metallicity vs. Galactocentric radius with a fit ($\pm$ 95\% confidence) for illustrative purposes. Data represent Cepheids, Quintuplet cluster LBVs and the Scutum Red Supergiant clusters \citep{Andrievsky:2002a, Andrievsky:2002b, Andrievsky:2002, Luck:2006, Pedicelli:2009}.}
  \label{fig:pw1}
  \vspace{-5pt}
\end{figure}

To first order, metallicity is known to increase towards the Galactic center. Recent studies of H II regions \citep{Balser:2011} and classical Cepheids \citep{Pedicelli:2009} define a clear gradient in metallicity in the Galactic disk (Figure \ref{fig:pw1}).  This gradient is traced by iron, as well as the $\alpha$-elements O, Ca, Si, Mg and Ti relative to H. However these gradients are slight, and measurements indicate that [$\alpha$/H] and [Fe/H] deviate from solar by little more than 0.5 dex as far out as 16 kpc from the Galactic center.

For the outer disk ($R_{\mathrm{GC}}$  $> 8$ kpc), [Fe/H] ratios in Cepheids increase with decreasing $R_{\mathrm{GC}}$ with a gradient of $\sim -0.05$ dex kpc$^{-1}$.  Between 8 and 4 kpc of the Galactic Center the [Fe/H] gradient is observed to be $\sim -0.02$ dex kpc$^{-1}$ with a maximum of $\sim$ 0.3 $\pm$ 0.1 dex at $R_{\mathrm{GC}} \sim$ 4 kpc (Figure \ref{fig:pw1}).  Inside R$_{GC} = $ 4 kpc, the [Fe/H] trend seems to ``roll over''.  Studies of Scutum Red Supergiant clusters at the end of the Galactic bar report sub-solar [Fe/H] ratios, with luminous blue variables (LBVs) and red supergiants (RSGs) in the Galactic center having observed values of [Fe/H] within error of the solar value \cite{Cunha:2007}. Measurements of oxygen and the $\alpha$-elements exhibit slightly more variability, with estimated maxima in [O/H] and [$\alpha$/H] at the Galactic center of ~ 0.5 dex and more typical values near 0.2 dex \citep{Najarro:2009, Davies:2009}. These results imply that the outer disk evolves somewhat differently than the inner disk and Galactic center.  \cite{Riquelme2010} used $[^{12}{\rm C}]/[^{13}{\rm C}]$ ratios to trace the infall of more chemically primitive gas in the halo  and the outer disk into the Galactic center region.  Their study illustrates the utility of Galactic Chemical Evolution of isotopes as a tracer of gas motions over time. 

\subsection{Galactic Chemical Evolution of Light Stable Isotopes}

Ratios of the stable isotopes of oxygen, carbon and nitrogen have been used as tracers of GCE.   Galactic chemical evolution leads to time dependent shifts in the isotopic makeup of the Galaxy, and this variability that results from the varying rates of astration and production should also be evident in variations with R$_{\rm GC}$. Isotope ratios have the advantage of normalizing some of the vagaries associated with production terms for the elements.  \citet{Tinsley:1975} provided a basis for a mathematical formalism to describe the GCE of nuclides.  In this treatment and those that followed, the rate of nuclide growth in the Galaxy is expressed as a function of both the star formation rate (SFR) within the Galaxy, $\Psi(t)$, and the initial mass function (IMF), $\phi(m)$, for the stellar sources.

\subsubsection{Primary Nuclides}

Nucleosynthetic processes requiring only primordial matter are termed primary processes, and produce \emph{primary} nuclides. Assuming that $M_{gas}(0) = M_{tot}$ and the mass of nuclide $i$ at time zero $M_i(0) = 0$, the equation for the evolution of the mass of a primary nuclide $p$ takes the form
\begin{equation} \label{eq:pnp1}
\frac{d}{dt}(M_{gas} X_p) = -\psi(t) X_p + E_p(t),
\end{equation}
where $\psi$ is the star formation rate, $X_p$ is the fractional abundance of nuclide $p$ in the ISM, $\psi(t) X_p$ is the rate of astration of nuclide $p$ due to new star formation and the ejection rate $E_p(t)$ is the rate at which both enriched and unenriched mass is returned to the ISM by supernovae and stellar winds.  The ejection rate can be written as  
\begin{equation} \label{eq:pnp2}
E_p(t) = \int_{m(t)}^{m_{u}} Y_p(m) \psi(t - \tau_m) \phi(m) \ dm,
\end{equation} 
where $m$ is the mass of a star with lifetime $\tau_m$, $Y_p(m)$ is the stellar yield of nuclide $p$ for a star of mass $m$, and $\psi(t - \tau_m)$ is the star formation rate at time of birth of the star of mass $m$. Integrating Equation (\ref{eq:pnp2}) over the chosen SFR and IMF yields an integro-differential equation which can be difficult to solve analytically.  For presentation purposes the simplifying assumption  that all stars with $m < m_{\sun}$ are immortal and all others die instantly is often made and is known as the ``instantaneous recycling approximation'' (IRA). By invoking the IRA and neglecting stellar lifetimes $\tau_m$, and using  the identity $d(M_{gas} X_p)/dt = M_{gas} dX_p/dt + X_p dM_{gas}/dt$,  Equation (\ref{eq:pnp1}) becomes
\begin{equation} \label{eq:pnp3}
\frac{dX_p}{dt} = \frac{1}{M_{\rm gas}} [(1 - R)\rho_p\psi(t) + f_{\rm in}(X_{p}' - X_{p}(t))],
\end{equation}
where $\rho_p$ is the IMF- integrated yield of new nuclides $p$  per unit stellar remnant mass, $R$ is the fraction of astrated material returned to the ISM, $(1-R)$ is the fraction of mass sequestered in stellar cores, $f_{in}$ is the flux of fresh gas to the Galaxy, and $X_p'$ is the abundance of nuclide $p$ for the infalling material.  In this expression $(1 - R)\rho_i\psi(t)$ is the mass of newly produced nuclide $p$ ejected from stars into the ISM per time.  Thus primary nuclide production is decoupled from stellar metallicity and is proportional to the star formation rate $\psi(t)$ and inversely proportional to the mass of gas remaining in the galaxy.  The solution to Equation (\ref{eq:pnp3}) for a simple closed box model where $f_{\rm in} =0$ is \citep{Searle:1972, Tinsley:1974, Prantzos:2008}  
\begin{equation} \label{eq:pnp4}
X_p(t) - X_p(0) = \rho_p \ \mathrm{ln}\left(\frac{M_{\rm tot}}{M_{\rm gas}}\right) = \rho_p \ \mathrm{ln}(\frac{1}{\mu_{\rm gas}}),
\end{equation}
where $\mu_{gas}$ is the fraction of total mass that is gas in the system.  A commonly used parameterization for the decrease in gas in the Galaxy with time is $\mu_{\rm gas} = \mu_{\rm gas}^o \exp{(-t/T)}$ where $T$ is a characteristic timescale that scales with the terminal age of the Galaxy.  We have in this case $X_p(t) - X_p(0)  = \rho_p(t/T)$ where $\mu_{\rm gas}$ is unity at $t=0$, showing that the amount of a primary nuclide grows roughly linearly with time. In what follows we set $X_p(0) = 0$ for convenience of presentation.

\subsubsection{Secondary Nuclides}

Odd-$Z$ and neutron-rich nuclides are often not accessible by way of primary nucleosynthetic processes, and production is dependent on the presence of primary``seed" nuclei synthesized in previous stellar generations.  In terms of IMF-integrated yields, $\rho_s = \alpha X_p$ where $\rho_s$ is the yield for the secondary nuclide and $\alpha$ is the proportionality constant relating secondary yield to primary seed  abundance.  The equation for the evolution of mass of a secondary nuclide $s$ is by analogy to  Equation (\ref{eq:pnp4})

\begin{equation} \label{eq:snp6}
X_s = \alpha X_p\ \mathrm{ln}\left(\frac{1}{\mu_{\rm gas}}\right) = \frac{\alpha }{\rho_p}X_p^2 .
\end{equation}
Since $X_p$, the fractional abundance of primary nuclide $p$, is expected to vary roughly linearly with time, Equation (\ref{eq:snp6}) shows that the abundance of the secondary nuclides should vary roughly as $t^2$ because  $X_s=\alpha \rho_p (t/T)^2$. 

The ratio of secondary to primary nuclides is 
\begin{equation} \label{eq:snp7}
\frac{X_s}{X_p} =\frac{\alpha X_p}{\rho_p}\propto Z,
\end{equation}
where $Z$ is the metallicity.  It follows that $X_s/X_p = \alpha (t/T)$ so the secondary-to-primary ratios should rise \emph{linearly} with time.  A valuable prediction is that the ratio of one secondary isotope to another will remain constant in this closed-system IRA treatment. 

The variation in molecular gas surface density across the Galaxy resembles the metallicity variation with $R_{\rm GC}$ shown in Figure \ref{fig:pw1} \citep{Heyer2015} in showing a monotonic increase moving inward from about 10 kpc to 5 kpc and a decrease from about 4 to 5 kpc toward the Galactic center.  This correspondence between metallicity and molecular gas surface density in the Milky Way suggests a link between time-averaged stellar processing and gas density, as suggested by the Schmidt-Kennicutt relationship between star formation rate and gas surface density \citep{Kennicutt1998, Kennicutt2012}.   As with overall metallicity $Z$, the abundances of primary nuclides of particular interest are also expected to vary with $R_{\rm GC}$.  We expect $\mu_{\rm gas}$ to decline towards the Galactic center in a closed system.  Comparisons between the sharp decline in the mass of stars with increasing $R_{\rm GC}$ \citep{Kent1991} and the more gradual declines in molecular and total gas surface densities with$R_{\rm GC}$ \citep{Heyer2015} show that $\mu_{\rm gas}$ does indeed decrease with smaller  $R_{\rm GC}$ in the Milky Way.  This is also the case for other, nearby spiral galaxies \citep{Leroy2008}.    For illustration purposes, a function for $\mu_{\rm gas}(R_{\rm GC}(kpc))$ with a range of $0$ to $1$  from the Galactic center to the outer Galactic disk  can be written as

\begin{equation} \label{eq:snp8}
\mu_{\rm gas} = 1-\frac{1}{R_{\rm GC}(kpc)+1}.
\end{equation} 
Substituting Equation (\ref{eq:snp8}) into Equation (\ref{eq:pnp4}) with $X_i(0)=0$ yields

\begin{equation} \label{eq:snp9}
X_p(t) =\rho_p \ \ln\left(\frac{R_{\rm GC}(kpc)+1}{R_{\rm GC}(kpc)}\right)
\end{equation}
which reduces to $X_p(t) \sim \rho_p/{R_{\rm GC}(kpc)}$ for $R_{\rm GC}>>1 {\rm kpc}$, showing that the relative abundances of primary nuclides should increase towards the Galactic center.   From Equations (\ref{eq:snp7}) and (\ref{eq:snp9}) we have that the ratio of secondary nuclides to primary nuclides should also vary inversely with $R_{\rm GC}$ since 

\begin{equation} \label{eq:ratio_v_RGC}
X_s/X_p\sim \alpha/R_{\rm GC}. 
\end{equation}
From these closed-system IRA equations dating back to Tinsley's early work, we have the basis for the expectation that at any given time in the Galaxy, secondary-to-primary isotope ratios should increase towards the Galactic center.  A corollary is that two distinct ratios, $X_{s,1} / X_p$ and $X_{s,2} / X_p$, composed of two distinct secondary nuclides and a single primary nuclide (e.g., $^{18}\mathrm{O} / ^{16}\mathrm{O}$ and $^{17}\mathrm{O} / ^{16}\mathrm{O}$ or $^{30}\mathrm{Si} / ^{28}\mathrm{Si}$ and $^{29}\mathrm{Si} / ^{28}\mathrm{Si}$) will tend to grow in lockstep.
The apparent chemical and isotopic ``age'' of the ISM should increase with decreasing $R_{\mathrm{GC}}$ in a manner that mimics the effects of time. For this reason, Galactocentric radius is in principle a proxy for time, and variations in isotope ratios with $R_{\rm GC}$ can be used as models for temporal variations in Galactic isotope abundance ratios. 

There are numerous mitigating factors that complicate the simple picture developed above.  Foremost among them is that the Galactic disk is not a closed system.  The effects of infalling gas towards the center of the Galaxy may be evidenced in  Figure \ref{fig:pw1} where metallicity is seen to level off or even decline near the Galactic center.  Despite these complicating factors, the prediction is that there should be a general relationship between metallicity and secondary/primary stable isotope ratios, and that the trend similar to that shown in Figure \ref{fig:pw1} should also obtain for these isotope ratios as well. If this prediction is verified, then we have good evidence that our understanding of the isotopic effects of GCE is reasonable, permitting us to extrapolate stable isotope ratios back in time, for example. Conversely, if a comparable trend is not observed, then we need to reconsider the significance of isotope ratio variations with $R_{\mathrm{GC}}$ and our ability to make inferences about the time evolution of stable isotope ratios.

\subsection{Previous Observations}

The secondary/primary isotopic abundance ratios of  oxygen \citep[e.g.,][]{Penzias:1981a,Wilson:1981} and carbon \citep[e.g.,][]{Langer:1990,Langer:1993,Milam:2005,Savage:2002,Wilson:1981,Wilson:1994} have have been used extensively to trace variations in the degree of astration across the Galaxy. 
$^{12}{\rm C}$ is produced during the helium-burning phase by the 3$\alpha$ reaction \citep{B2FH:1957} and is the second most abundant non-primordial nuclide \citep{Clayton:2003}. While $^{12}$C is a primary nucleosynthetic product, $^{13}$C is a secondary nucleosynthetic product, requiring pre-existing $^{12}$C for efficient production \citep{B2FH:1957}. Approximately half of the carbon in the ISM originates from Type II supernovae, while the remainder is produced by intermediate mass (1.5 - 6 M$_{\sun}$) asymptotic giant branch (AGB) stars \citep{Clayton:2003}.   \citet{Milam:2005} showed that the $[^{13}{\rm C}]/[^{12}{\rm C}]$ ratios\footnote{Brackets are used to distinguish atomic abundances from mass abundances, but [x]/[y] should not be confused with [x/y] where only the latter is in dex units} in Galactic molecular clouds  increase towards the Galactic center, consistent with the qualitative expectations of GCE.  Based on this agreement between data and GCE predictions, the authors suggested that the higher $^{13}{\rm C}/^{12}{\rm C}$ in the ISM today relative to solar could be the consequence of $^{13}{\rm C}$ enrichment relative to $^{12}{\rm C}$ over the last 4.6 Gyrs. 

The oxygen isotope system differs from the carbon system in that it has two stable heavy isotopes, $^{17}$O and $^{18}$O. The most abundant isotope of oxygen, $^{16}$O, is a primary nuclide produced during He burning. The rare isotopes, $^{17}$O, and $^{18}$O, are secondary nuclides. $^{17}$O is the daughter product of $^{17}$F, which undergoes rapid $\beta^-$ decay after being produced as part of the CNO tricycle. The preponderance of $^ {18}$O is produced by  $\alpha$ addition to $^{14}$N, which is in turn produced from $^{12}$C during the CNO tricycle. $^{18}$O is also produced from $^{17}$O \citep{Clayton:2003,B2FH:1957}.

\begin{figure}[b!]
  \begin{center}
    \includegraphics[width=0.47\textwidth]{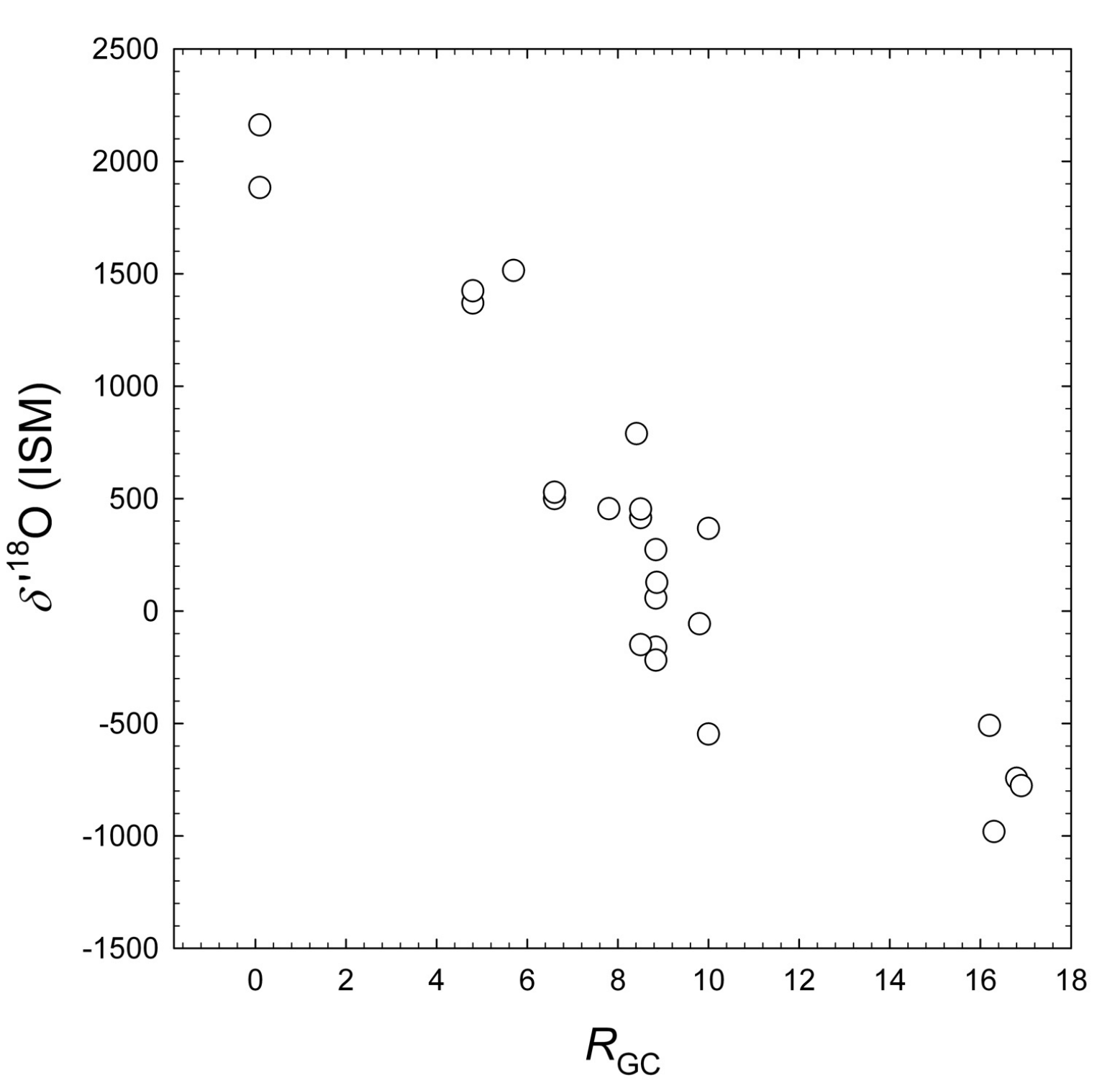}
  \end{center}
  \vspace{-10pt}
  \caption{[$^{18}$O]/[$^{16}$O] as $\delta^{\prime 18}$O in permil vs Galactocentric radius. References in text.}
  \label{fig:ovrgc}
  \vspace{0 pt}
\end{figure}

The existence of two secondary isotopes makes the oxygen system particularly attractive for tracing GCE.  Optical depth effects have hampered efforts to determine C$^{16}$O column densities within sources. However, one can use estimates for the [$^{12}$C]/[$^{13}$C] ratio within the source to calculate the C$^{16}$O column density from $^{13}$C$^{16}$O observations. Using this approach, Galactic oxygen isotope abundances can be extrapolated from the $^{13}$CO, C$^{18}$O, and C$^{17}$O column densities reported by \citet{Wouterloot:2008} and the [$^{12}$CO]/[$^{13}$CO] vs. $R_{\mathrm{GC}}$ data from \citet{Milam:2005}. For this and other purposes in this paper, we use the $\delta'$ notation commonly used in cosmochemistry to compare isotope ratios expressed as permil differences from a reference ratio such that $\delta^{\prime j}{\rm X} = 10^3\ln(R/R_{\rm ref})$, $R$ is the isotope ratio $[^jX]/[^iX]$, and $i$ and $j$ are the heavy and corresponding light isotopes, respectively (we use the logarithmic form of the $\delta$ notation to accommodate the large variations in isotope ratios across the Galaxy). The resulting [$^{18}$O]/[$^{16}$O] ratios, normalized to the reference ISM value of Wilson (1999), vs. $R_{\mathrm{GC}}$ is shown in Figure \ref{fig:ovrgc}.  These extrapolated data indicate that  [$^{18}$O]/[$^{16}$O] ratios increase linearly with decreasing $R_{\mathrm{GC}}$, in qualitative agreement with the predictions of secondary/primary increases with GCE. However, the range in  [$^{18}$O]/[$^{16}$O] of greater than a factor of 10, or $>900\%$ (a factor of 10 corresponds to 2300 per mil on the ordinate in Figure \ref{fig:ovrgc}, exceeds the theoretical predictions of \citet{Prantzos:1996}  by a factor of $\sim$ 2 to 3  \citep{Young:2011}  and appears to extend unabated into the Galactic center, contrasting with the "downturn" seen in both the [O/H] and [Fe/H] trends. 

\begin{figure}[t!]
  \begin{center}
    \includegraphics[width=0.47\textwidth]{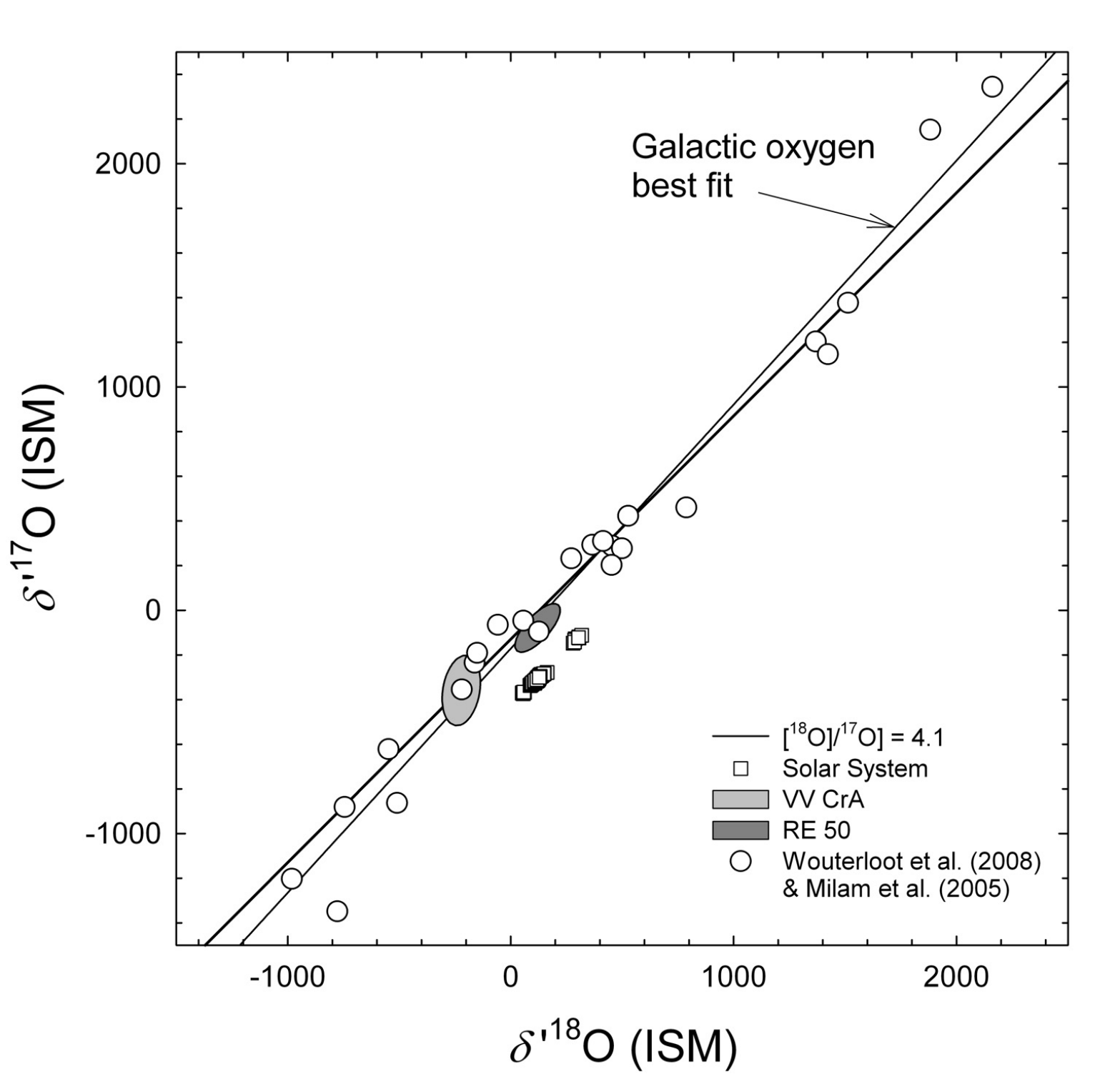}
  \end{center}
  \vspace{-10pt}
  \caption{Oxygen triple-isotope plot comparing molecular clouds obtained by a combination of radio observations (circles), young stellar objects obtained by IR absorption (error ellipses), and the solar system (squares).  The 1:1 line is compared with the best fit line with a slope of 1.11 $\pm 0.08$ $2\sigma$.  References are discussed in the text.}
  \label{fig:cotriple}
  \vspace{0pt}
\end{figure}

The two oxygen secondary/primary isotope ratios can also be used in concert to evaluate the presence or absence of GCE in the oxygen isotopologue data. On a so-called three-isotope plot in which [$^{17}$O]/[$^{16}$O] is plotted against [$^{18}$O]/[$^{16}$O], both normalized to a suitable reference, the first-order prediction based on Galactic chemical evolution is that data representing a range of localities across the Galaxy will define a slope of unity. Quantitative GCE models for the oxygen isotopes are in general agreement with the simplified equations presented above and show that even as [$^{17}$O]/[$^{16}$O] and [$^{18}$O]/[$^{16}$O]  have risen with time, the ratio of the two secondary nuclides, [$^{18}$O]/[$^{17}$O], should have been constant after the first billion years \citep{Timmes:1995,Prantzos:1996}. This is because both secondary nuclides have a similar dependency on metallicity in these models. Figure \ref{fig:cotriple} \citep[after ][]{Young:2011} illustrates that the [$^{17}$O]/[$^{16}$O] and [$^{18}$O]/[$^{16}$O] ratios across the galaxy define a slope in triple-isotope space of 1.11 $\pm$ 0.08 (2$\sigma$) that is practically indistinguishable from the unity value predicted by closed-system IRA GCE.  Also shown in Figure \ref{fig:cotriple} are infrared absorption data for young stellar objects that show less of a spread in oxygen isotope ratios, albeit in part because they are from sources near the solar circle. 

The validity of the combined carbon/oxygen data sets has been questioned on the basis that there is good reason to believe that $^{17}$O is produced mainly in intermediate mass stars  \citep{Romano:2003} while $^{18}$O is produced in more massive stars. In this case the progenitors of $^{17}$O live longer than those of $^{18}$O, allowing for deviations from expectations of nearly constant $^{18}$O/$^{17}$O with time (in effect altering $\alpha$  for the two secondary nuclides in Equation \ref{eq:ratio_v_RGC}). \citet{Nittler:2012} also question the veracity of the $\delta'^{18}$O vs $R_{\mathrm{GC}}$ trend, referring to "chemical" rather than isotopic partitioning to account for varying [$^{13}$C$^{16}$O]/[$^{12}$C$^{18}$O]. We point out that both the spatial and spectral resolution in the previous studies limited the ability to detect optical depth effects that would spuriously enhance the recovered [$^{18}$O]/[$^{16}$O]  and [$^{17}$O]/[$^{16}$O] ratios, artificially translating any affected sources up a slope-1 trajectory in triple-isotope space. Additionally, new $^{12}$C nuclei produced in the He-burning shells of AGB stars are ultimately conveyed to the outer envelopes of the stars during convective instability dredge-up events.  Consequently, a considerable amount of what is effectively primary $^{13}$C nuclei is created in the He intershell, some of which is then convectively transported to the surface and shed in stellar winds \citep{Gallino:1998,Straniero:1997}. The degree to which this effect biases Galactic carbon isotope ratios is not well quantified, and complicates the interpretation of these isotopes in the ISM.

For these reasons, the oxygen trends in Figures (\ref{fig:ovrgc}) and (\ref{fig:cotriple}) might be questioned. The trend seen in oxygen is commonly regarded as evidence for Galactic chemical evolution of the oxygen isotopes \citep{Wilson:1999}.

\section{Testing GCE using Silicon}

While interstellar oxygen isotopes have been extensively studied \citep[e.g.]{Wilson:1999}, the same is not true of the other light-element systems having 3 stable-isotopes; $^{24,25,26}$Mg and $^{28,29,30}$Si. Magnesium is poorly suited to widespread interstellar observations, however silicon is readily observed in molecular clouds at millimeter wavelengths. 

A number of silicon-bearing molecular species, including SiC, SiS, SiCN and SiNC, have been detected in the circumstellar envelopes of AGB stars, however the possibility of local nucleosynthesis makes these unsuitable proxies for the average interstellar abundances. SiO is commonly observed to trace shocks in dense, turbulent cloud cores and molecular outflows \citep{Ziurys:1989, Martin:2009, Caselli:1997, Schilke:1997} where it is thought to dominate the gaseous silicon budget and the chances that observational measurements are not representative of the bulk silicon composition are minimized. For this reason, SiO is well suited for probing isotopic GCE. Because silicon is a relatively refractory element and is largely sequestered in silicate dust, SiO column densities are typically modest in comparison to common molecules, such as CO, CS, or HCN, and observed SiO line intensities are similarly modest.  As a consequence of requiring relatively dynamic physical conditions, most sources of SiO emission are compact and efficient observation requires large telescopes to achieve favorable beam-filling factors. Fortunately, $^{29}$Si and $^{30}$Si are relatively abundant (with solar [$^{28}$Si]/[$^{29}$Si] = 19.7 and [$^{28}$Si]/[$^{30}$Si] = 29.8), allowing the weaker isotopologue lines to be accurately measured with feasible integration times. 

The silicon isotope system is largely analogous to that of oxygen, in that it contains one primary and two secondary nuclides. The primary silicon isotope, $^{28}$Si, is an alpha process nuclide and is by far the most prevalent, with a solar abundance of 92.23\% \citep{Clayton:2003}. $^{29}$Si and $^{30}$Si are both secondary, forming largely from $^{25}$Mg and $^{26}$Mg during Ne-burning, as well as during core-collapse Type II supernovae. Both rare isotopes also form from $^{28}$Si in the He-burning shells of AGB stars.  While contributions from He-burning AGB stars could alter local compositions, it likely has little effect on the overall isotopic budget of the interstellar medium (ISM) \citep{Clayton:2003}. GCE models  predict that, to first order, the silicon and oxygen isotope ratios should evolve in parallel.  Therefore, based on the oxygen data (e.g., Figures \ref{fig:ovrgc} and \ref{fig:cotriple}), one expects nearly constant [$^{29}$Si]/[$^{30}$Si] across the Galaxy, as well as radial gradients in the [$^{29}$Si]/[$^{28}$Si] and [$^{30}$Si]/[$^{28}$Si] ratios that increase with decreasing $R_{\mathrm{GC}}$.

\begin{figure} [t!]
  \begin{center}
    \includegraphics[width=0.5\textwidth]{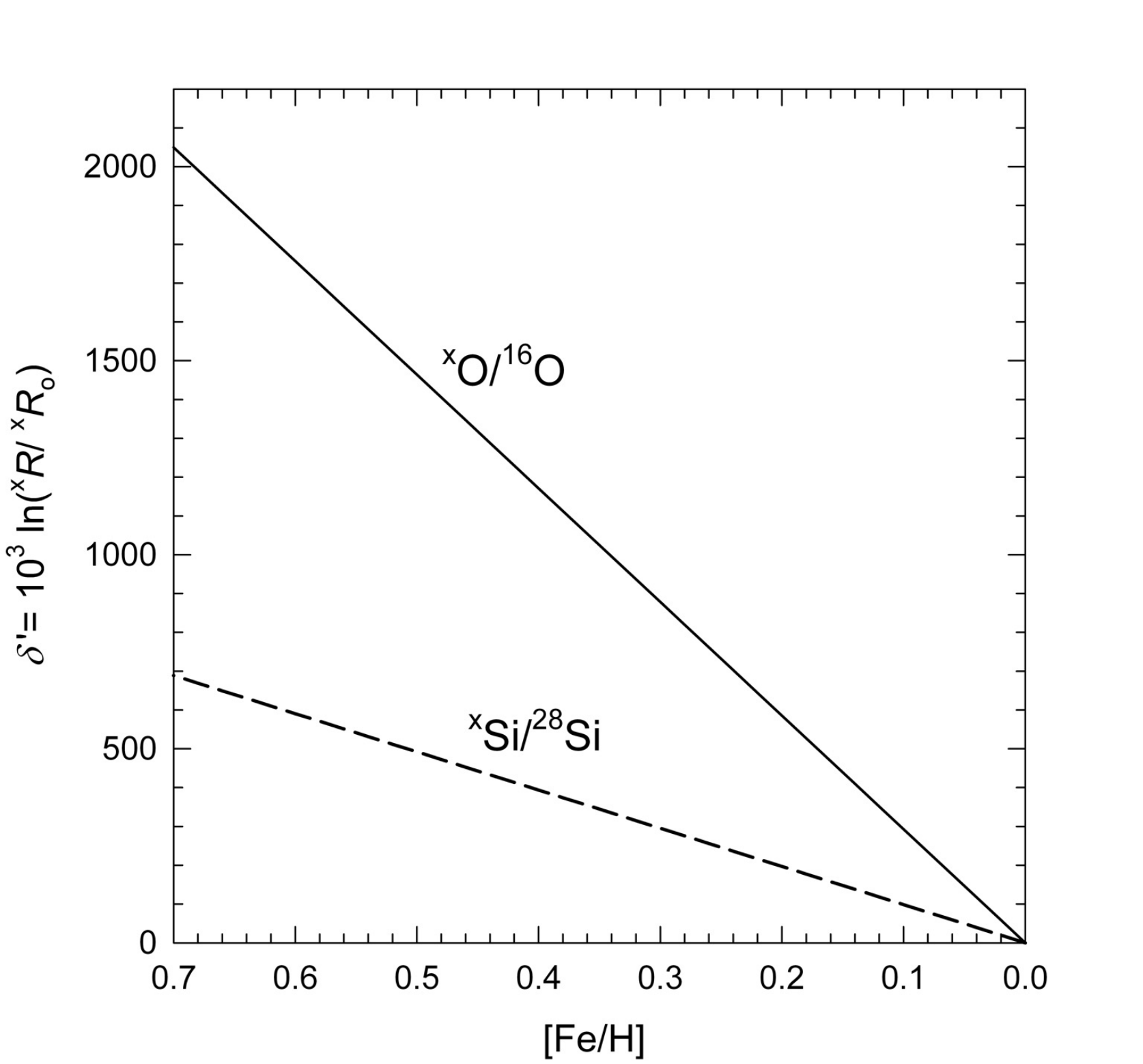}
  \end{center}
  \vspace{-10pt}
  \caption{Predicted dependence of oxygen and silicon isotope abundance ratios on local stellar [Fe/H] \citep{Timmes:1996,Timmes:1995} relative to solar. Secondary to primary isotopic ratio values on the ordinate expressed as $\delta' = 10^3 \ln({R}/{R}_o)$ where $R$ refers to Galactic values, and R$_o$ to the initial reference value.}
  \label{fig:co1}
  \vspace{0pt}
\end{figure}

Predictions for the magnitude of the variations in [$^{29}$Si]/[$^{28}$Si] and [$^{30}$Si]/[$^{28}$Si] relative to the variations in the oxygen isotope system can be made using the silicon isotope GCE model of \citet{Timmes:1996} and the oxygen isotope GCE model of \citet{Timmes:1995} (the Timmes and Clayton results are typical of numerous other models for [Fe/H] $\ge$ solar, \citealt{Lewis:2013}). The predicted dependencies of isotope ratios on metallicity are $d[^{j}{\rm Si}/^{28}{\rm Si}]/d[{\rm Fe}/{\rm H}] = 0.43$ and $d[^{j}{\rm O}/^{16}{\rm O}]/d[{\rm Fe}/{\rm H}] = 1.27$ where $j$ represents the heavy isotopes and all ratios are in dex. If the Galactic center is no greater than $\sim$ 0.5 dex in [Fe/H], as suggested by the observed metallicities of Quintuplet cluster LBVs \citep{Cunha:2007}, then one predicts an increase in [$^{18}$O]/[$^{16}$O]  expressed as $\delta^{\prime 18}$O relative to solar of approximately 1500 permil between the solar circle and the Galactic center. The corresponding increase in [$^{29}$Si]/[$^{28}$Si]  expressed as $\delta^{\prime 29}$Si is predicted to be $\sim$ 500 permil (Fig. \ref{fig:co1}). As described above, this prediction is similar to, but approximately $3 \times$ smaller than, the observed variation for the oxygen isotopes \citep{Wilson:1999,Young:2011}.

\begin{figure} [t!]
  \begin{center}
    \includegraphics[width=0.48\textwidth]{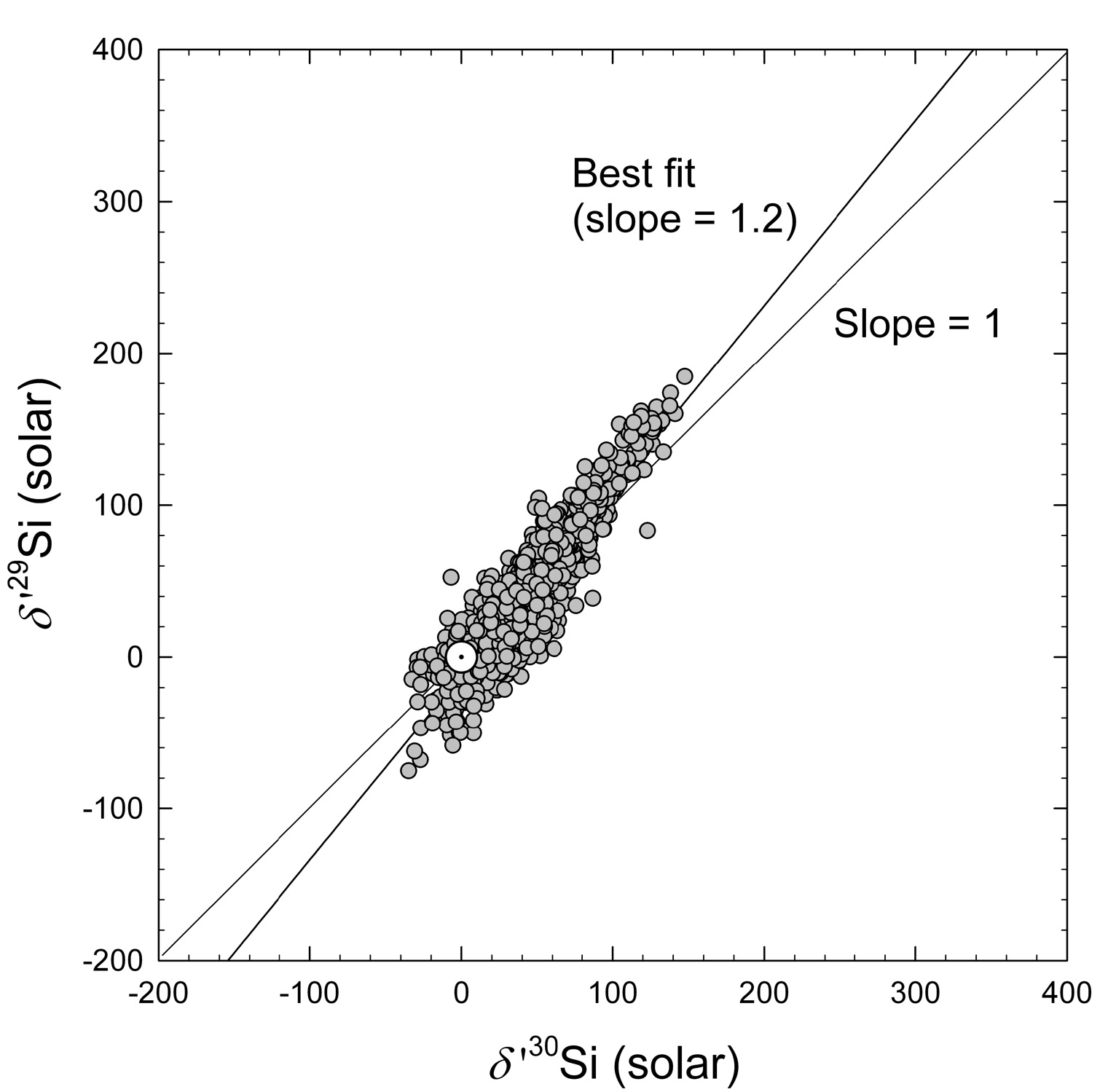}
  \end{center}
  \vspace{-10pt}
  \caption{Silicon isotope ratios of mainstream presolar SiC grains (grey filled circles) expressed as per mil deviations from the solar ratios, or  $\delta^{\prime 29}{\rm Si}$ vs. $\delta^{\prime 30}{\rm Si}$ relative to solar (data from Ernst Zinner, pers. comm.).  The white circle with the center dot indicates present-day solar abundances and defines the origin. The best-fit line has a slope of 1.22 $\pm 0.02$ $2\sigma$.}
  \label{fig:co2}
  \vspace{0pt}
\end{figure}

Additional motivation for establishing the Galactic distribution of silicon isotopes can be garnered from the [$^{29}$Si]/[$^{28}$Si] and [$^{30}$Si]/[$^{28}$Si] isotope abundance ratios found in presolar SiC grains. These grains predate the Sun and are thought to have condensed out of the winds expelled from ancient asymptotic giant branch (AGB) stars. The mainstream SiC grains ($> 90\%$ of all presolar SiC grains) define a spread in $[^{29}$Si]/[$^{28}$Si] (as $\delta^{\prime 29}{\rm Si}$) and [$^{30}$Si]/[$^{28}$Si] (as $\delta^{\prime 30}{\rm Si}$) along a slope of $\sim 1.2$ (Figure \ref{fig:co2}). The variation in silicon isotope ratios is an order of magnitude larger than that expected from  nucleosynthesis in a single AGB star and it is generally agreed that the spread in [$^{29}$Si]/[$^{28}$Si] and [$^{30}$Si]/[$^{28}$Si] predates the AGB parents of these grains \citep[and references therein]{Lugaro:1999}. 

The considerable spread in the presolar SiC [$^{29}$Si]/[$^{28}$Si] and [$^{30}$Si]/[$^{28}$Si] ratios represents either a manifestation of GCE as sampled by AGB stars with different birth dates, or heterogeneity in the ISM material from which the AGB stars formed. GCE predicts that solar [$^{29}$Si]/[$^{28}$Si] and [$^{30}$Si]/[$^{28}$Si] ratios representing the ISM when the Sun formed  4.6 Gyr before present should be larger than the [$^{29}$Si]/[$^{28}$Si] and [$^{30}$Si]/[$^{28}$Si] ratios found in presolar SiC grains that predate the Sun, but this is not observed. This apparent excess in $^{28}$Si (or depletion in $^{29}$Si and $^{30}$Si) in the Sun is a conundrum. \citet{Alexander:1999} suggested that the solar system was enriched in $^{28}{\rm Si}$ by supernova ejecta.  A model for that enrichment was given by \citet{Young:2011}.  Alternatively, \citet{Lugaro:1999} suggested that the distribution of data in Figure \ref{fig:co2} can be explained simply by dispersion resulting from incomplete mixing of stellar sources, although this model fails to reproduce correlations between Ti and Si isotope ratios in the SiC grains \citep{Nittler:2005}. More recently, \citet{Lewis:2013} used the SiC grain data and GCE models to derive the metallicity [Fe/H] and ages of the SiC parent stars. Their results suggest a distribution in [Fe/H] with a mean near solar and a $1\sigma$ error of about $0.2$ dex with a skew towards higher [Fe/H]. Their derived range in metallicity is less than that observed in the solar neighborhood today. Mapping the distribution of Galactic Si isotope ratios as a function of $R_{\rm GC}$ will provide much needed context for the questions raised by the comparison between solar  and presolar SiC grain Si isotope ratios. 

\section{Observations}

Initial observations of the $v=0$, $J=1 \to 0$ transition of the three silicon isotopologues of SiO were carried out at the Robert C. Byrd Green Bank radio telescope  (GBT) in May of 2013 (project GBT13A-415).  Additionally, several weeks were spent in  Green Bank in January and February of 2014 making follow-up observations (project GBT14A-431). Seven sources with known radial distances from the Galactic center and brightness temperatures between 1 and 3 Kelvin were selected, including GCM-0.13-0.08 ($R_{\rm GC}~\lesssim$ 0.1 kpc), GCM0.11-0.11 ($R_{\rm GC}~\lesssim$ 0.1 kpc), W51e2 ($R_{\rm GC}$ = 6.4 kpc), DR21(OH) ($R_{\rm GC}$ = 7.9 kpc), L1157 ($R_{\rm GC}$ = 8.1 kpc), NGC 7538 S ($R_{\rm GC}$ = 9.3 kpc), and AFGL 5142 ($R_{\rm GC}$ = 9.8 kpc) (Table 1).  Because of the compact nature of many of the sources in this study, we used main beam temperatures ($T_{\rm mb}$) rather than antenna temperatures (Table 1). 

Excluding the two in the Galactic center, all of the  sources can be described broadly as SiO produced by shock-induced evaporation of silicate grains associated with protostellar outflows in sites of massive star formation.  AGFL 5142 is a cluster of high-mass protostars \citep{Zhang2007}.  DR21(OH) is a site of dense molecular clouds within Cygnus X where several OB stars are resident \citep{Duarte2014}.  L1157 is a dark cloud in Cepheus harboring young protostars with chemically active outflows \citep{Nisini:2007}. NGC 7538 S is a high-mass accretion disk candidate comprising a compact H II region surrounding a nascent O star in the Perseus spiral arm \citep{Naranjo2012}.  W51e2 is a bright ultracompact H II region in the W51 star-forming region. Hints of bipolar outflows perpendicular to a rotating ionized disk are reported, as is evidence for a newly formed O star or cluster of B stars \citep{Shi2010}.  SiO in the Galactic center traces shocked high-velocity molecular cloud gas there.  GCM $-$0.13$-$0.08 is also known as the 20 km/s cloud and is one of the densest clouds in the Sagittarius A (Sgr A) cloud complex \citep{Tsuboi2011}.  GCM0.11$-$0.11 is another member of the Sgr A cloud complex that appears to be composed of a composite of hot and dense clumps \citep{Handa2006}.
 
Data for all three isotopologues of SiO were collected simultaneously using the Q-band receiver and autocorrelation spectrometer backend. The autocorrelation spectrometer accommodated four spectral windows, one for each of the three silicon isotopologues of SiO, and a 'spare'  that was put to use in several capacities that will be addressed in subsequent sections. The two Galactic center sources were observed using 200 MHz bandpass windows with 24.4 kHz wide channels yielding $\sim$ 340 m/s resolution, and all other sources utilized  50 MHz bandpass windows with 6.1 kHz channels yielding $\sim$ 85 m/s resolution. These spectral resolutions translate to resolving powers of $\sim 8.8\times10^5$ and $\sim 3.5\times10^6$ respectively;  the emission lines from all sources are well resolved. 

Pointing was checked against nearby 7 mm continuum sources every hour, and errors were typically 3 arcseconds or less. All observations were made using in-band frequency switching, and all switching was by $40\%$ of the bandpass at a rate of 2 Hz. System temperatures hovered around $\sim$ 80 K for most observations, but varied from lows of about 70K to highs of 130K at low elevations or in inclement weather. We found that most sources required approximately 3 hours of integration time to achieve the desired signal-to-noise ratio for the rare emission line. Noise temperatures (prior to resampling) on the order of 20mK were achieved in most sources.

\begin{deluxetable*}{lccccrrrr}
\tabletypesize{\footnotesize} 
\tablecolumns{9} 
\tablecaption{List of sources and observed SiO v=0, J=$1\to0$ emission lines. }
\tablehead{\colhead{Source} & \colhead{$\alpha$} & \colhead{$\beta$} & \colhead{Pointing Offset} & \colhead{Species} & \colhead{$T_{\mathrm{mb}}$} & \colhead{$\Delta v_{1/2}$}  & \colhead{$\int T_{\mathrm{mb}} \, \mathrm{d}v$}  & \colhead{$V_{\mathrm{LSR}}$} \\ 
\colhead{} & \colhead{(J2000)} & \colhead{(J2000)} & \colhead{(",")} & \colhead{} & \colhead{(K)} & \colhead{($\mathrm{km \, s}^{-1}$)} & \colhead{($\mathrm{K \, km \,  s}^{-1}$)} & \colhead{($\mathrm{km \, s}^{-1}$)} } 
%% All data must appear between the \startdata and \enddata commands
\startdata
DR21 (OH)  &  20:39:01.0  &  +42:22:50  &  (0, -5)  &  $^{28}$SiO  &  1.484 $\pm$ 0.019  &  5.14 $\pm$ 0.03  &  9.626 $\pm$ 0.025  & -4.69 $\pm$ 0.17  \\
&                  &                &              &  $^{29}$SiO  &  0.081 $\pm$ 0.016  &  5.40 $\pm$ 0.29  &  0.529 $\pm$ 0.016  &  -4.62 $\pm$ 0.17 \\
&                  &                &              &  $^{30}$SiO  &  0.057 $\pm$ 0.017  &  5.02 $\pm$ 0.41  &  0.348 $\pm$ 0.016  &  -4.46 $\pm$ 0.17 \\
\\
L1157 B1  &  20:39:06.4  &  +68:02:13  &  (0, 0)  &  $^{28}$SiO  &  3.376 $\pm$ 0.019  &  3.63 $\pm$ 0.01  &  14.080 $\pm$ 0.016  & 1.80 $\pm$ 0.17 \\
&                 &               &               &  $^{29}$SiO  &  0.280 $\pm$ 0.016  &  3.19 $\pm$ 0.08  &  1.074 $\pm$ 0.013  & 1.87 $\pm$ 0.17 \\
&                 &               &               &  $^{30}$SiO  &  0.189 $\pm$ 0.016  &  3.27 $\pm$ 0.10  &  0.703 $\pm$ 0.013  & 1.76 $\pm$ 0.17 \\
\\
NGC 7538 S  &  23:13:44.8  &  +61:26:51  &  (0, -5)  &  $^{28}$SiO  &  1.783 $\pm$ 0.022  &  4.83 $\pm$ 0.03  &  11.823 $\pm$ 0.043  &  -54.22 $\pm$ 0.17 \\
 &                &               &                &  $^{29}$SiO   &  0.118 $\pm$ 0.020  &  4.58 $\pm$ 0.18  &  0.729 $\pm$ 0.023  &  -54.21 $\pm$ 0.17 \\
 &                &               &                &  $^{30}$SiO   &  0.089 $\pm$ 0.020  &  4.41 $\pm$ 0.25  &  0.522 $\pm$ 0.023  &  -54.19 $\pm$ 0.17 \\
\\
AFGL 5142  &  05:30:45.9  &  +33:47:56  &  (+25, -5)  &  $^{28}$SiO  &  0.920 $\pm$ 0.013  &  5.93 $\pm$ 0.04  &  7.467 $\pm$ 0.037  &  -2.71 $\pm$ 0.17 \\
 &                &               &                &  $^{29}$SiO   &  0.052 $\pm$ 0.013  &  5.74 $\pm$ 0.57  &  0.401 $\pm$ 0.016  &  -2.41 $\pm$ 0.17 \\
 &                &               &                &  $^{30}$SiO   &  0.035 $\pm$ 0.012  &  6.12 $\pm$ 0.72  &  0.289 $\pm$ 0.015  &  -1.88 $\pm$ 0.17 \\
\\
W51e2  &  19:23:42.0  &  +14:30:00  &  (+25, +30)  &  $^{28}$SiO  &  2.257 $\pm$ 0.014  &  8.22 $\pm$ 0.02  &  21.620 $\pm$ 0.042  &  -56.30 $\pm$ 0.17 \\
 &                &               &                &  $^{29}$SiO  &  0.142 $\pm$ 0.011  &  7.63 $\pm$ 0.19  &  1.264 $\pm$ 0.013  &  -56.33 $\pm$ 0.17 \\
 &                &               &                &  $^{30}$SiO  &  0.094 $\pm$ 0.012  &  8.35 $\pm$ 0.24  &  0.856 $\pm$ 0.014  &  -55.98 $\pm$ 0.17 \\
\\
GCM0.11-0.11  &  17:46:18.0  &  -28:54:00  &  (+40, +35)  &  $^{28}$SiO  &  1.791 $\pm$ 0.026  &  19.38 $\pm$ 0.14  &  43.685 $\pm$ 0.411  &  -23.37 $\pm$ 0.67 \\
 &                  &                 &                  &  $^{29}$SiO  &  0.168 $\pm$ 0.031  &  16.37 $\pm$ 2.48  &  3.339 $\pm$ 0.117  &  -22.66 $\pm$ 0.67 \\
 &                  &                 &                  &  $^{30}$SiO  &  0.106 $\pm$ 0.020  &  17.09 $\pm$ 1.01  &  2.240 $\pm$ 0.111  &  -23.67 $\pm$ 0.67 \\
\\
GCM-0.13-0.08  &  17:45:25.2  &  -29:05:30  &  (+180, +70)  &  $^{28}$SiO  &  4.195 $\pm$ 0.028  &  19.88 $\pm$ 0.04  &  91.133 $\pm$ 0.510  &  -17.25 $\pm$ 0.67 \\
 &                  &                 &                    &  $^{29}$SiO  &  0.361 $\pm$ 0.036  &  17.45 $\pm$ 0.43  &  6.950 $\pm$ 0.129  &  -16.54 $\pm$ 0.67 \\
 &                  &                 &                    &  $^{30}$SiO  &  0.250 $\pm$ 0.024  &  17.13 $\pm$ 0.41  &  4.532 $\pm$ 0.090  &  -17.36 $\pm$ 0.67 \\
\enddata
\end{deluxetable*}

\section{Calibration and Data Reduction}

The calibration and reduction of all data reported here  were done using a novel suite of IDL and Fortran programs (the HYDRA software package) written by one of us (NNM) and verified by consultation with GBT staff astronomers (these functions expand upon the basic data reduction afforded by the GBTIDL software package).  The procedures include several vectorized approaches to the calibrations that enhance accuracy and precision of the extracted line profiles. 

\subsection{Flux Calibration}

As a consequence of the sensitive nature of the measurements being made, special attention was paid to flux calibration to ensure that any drift in receiver performance between observations could be corrected. Differences in receiver gain between spectral windows were also of special concern.

The primary concern with the standard approach for calculating system temperatures, $T_{\rm sys}$,  and calibration temperatures,  $T_{\rm cal}$, is that any information about frequency-dependent gain \emph{within} the bandpass is lost. Although atmospheric opacity and aperture efficiency are largely invariant across 50 MHz and 200 MHz spectral windows, noise diode power output and LO/IF system response are not. Left unaccounted for, these frequency dependencies are an unacceptably large source of potential error. In order to mitigate these effects, the standard calibration protocol has been adapted to account for channel-by-channel variations in the system response by substituting array valued, or "vectorized', versions of calibration and system temperatures,  $\vec{T}_{\rm cal}$ and $\vec{T}_{\rm sys}$, for their standard scalar valued counterparts. Vectorized calibration routines were developed expressly for this survey as part of the HYDRA data pipeline, allowing gain profiles to be determined pixel-by-pixel across the entire bandpass, thereby accommodating any frequency dependence that may be present. Further, gain profiles for each IF, polarization, noise diode state and frequency position were calculated independently to ensure uniform calibration. 

The GBT Q-band receiver was calibrated using a noise diode integrated into the primary signal path. The diode was calibrated against nearby radio-loud active galactic nuclei, 3C405, 3C286 or 3C147, at the beginning and end of each observing period. The spectral flux density of the calibrator, $\vec{S}_{\rm source}$, was calculated using the polynomial expression and coefficients reported by Perley \& Butler (2012) and converted to a source main beam temperature, $\vec{T}_{\rm source}$, with the expression
\begin{equation}
\vec{T}_{\rm source} = 2.85\, \vec{S}_{\rm source}\  \frac{\vec{\eta}_{\rm a}}{\vec{\eta}_{\rm mb}}\exp\left(\frac{-\vec{\tau}_o}{\sin{({\theta})}}\right),
\end{equation}
where $2.85 = (A_g / 2k_b)$ is the GBT-specific gain constant defined by the physical collecting area $A_g$,  $\vec{\tau}_o$ is the zenith atmospheric opacity  estimated from $\vec{\tau}_o = 0.008 + \exp(\sqrt{\vec{\nu}}\,)/8000$ where $\nu$ is the frequency in GHz,  $\vec{\eta}_{\rm a}$ is aperture efficiency, $\vec{\eta_{\rm mb}}$ is main beam efficiency (1.37 times $\vec{\eta}_{\rm a}$), and $\theta$ is the altitude.  Aperture efficiency is  estimated using Ruze's equation (Ruze, 1952; Ruze, 1966) with the GBT-specific peak aperture efficiency of $0.71$ and RMS surface accuracy of $390$ microns.  The calculated source temperature was then used to convert the power output of the noise diode to a calibration temperature profile:
\begin{equation}\label{eq:calib}
\vec{T}_{\rm cal} = \vec{T}_{\rm source} \left[\frac{\rm Src^{on} - Src^{off} + Sky^{on} - Sky^{off}}{\rm Src^{on} - Sky^{on} + Src^{off} - Sky^{off}}\right].
\end{equation}
In Equation (\ref{eq:calib}) "Src" and "Sky" refer to the source and sky positions and superscripts ``on'' and ''off'' refer to the state of the noise diode.  Calibration temperatures are obtained for each polarization and frequency position. The flux calibrators were observed for either two or four 30 second integrations followed by an equal number of sky integrations offset by $-0.5$ degrees in azimuth, and the noise diode calibration temperature for each polarization and frequency position was independently calculated for each of the either four or sixteen possible Src/Sky integration pairs. 

System temperature profiles were found to differ somewhat between frequency positions. To account for this, all spectra were folded using a channel-by-channel weighted mean, where the weight of each channel is equal to the inverse square of the system temperature in that channel, such that the main beam temperature is
\begin{equation}\label{eq:mean}
\vec{T}_{\rm mb}(n) = \frac{\vec{T}^{\rm sig}(n)[\vec{T}_{\rm sys}^{\rm sig}(n)]^{-2} + \vec{T}^{\rm ref}(n)[\vec{T}_{\rm sys}^{\rm ref}(n)]^{-2}}{[\vec{T}_{\rm sys}^{\rm ref}(n)]^{-2} + [\vec{T}_{\rm sys}^{\rm sig}(n)]^{-2}},
\end{equation}
where $n$ is the channel index and the ``sig'' and ``ref'' superscripts refer to the signal and reference frequency positions, respectively. All subsequent averaging operations between polarizations, integrations, scans and observations were done using the same channel-by-channel weighted mean.   

\subsection{Baselines}

The vectorized calibration routine tamed the baselines but did not eliminate all structure. Typical low frequency ($\nu \sim $  bandpass) baselines were fit with low-order polynomials for subtraction. However, differentiating between baseline structure and emission-line structure was challenging in the low-brightness sources DR21(OH) and AFGL 5142.  In order to avoid confusing line wings with baselines, we omitted all velocities from the baseline fits that lay within  $\pm 3$ times the full-width half-maximum (FWHM) of the $\rm ^{28}SiO$ line. 

\begin{figure*} [t]
  \begin{center}
    \includegraphics[width=0.90\textwidth]{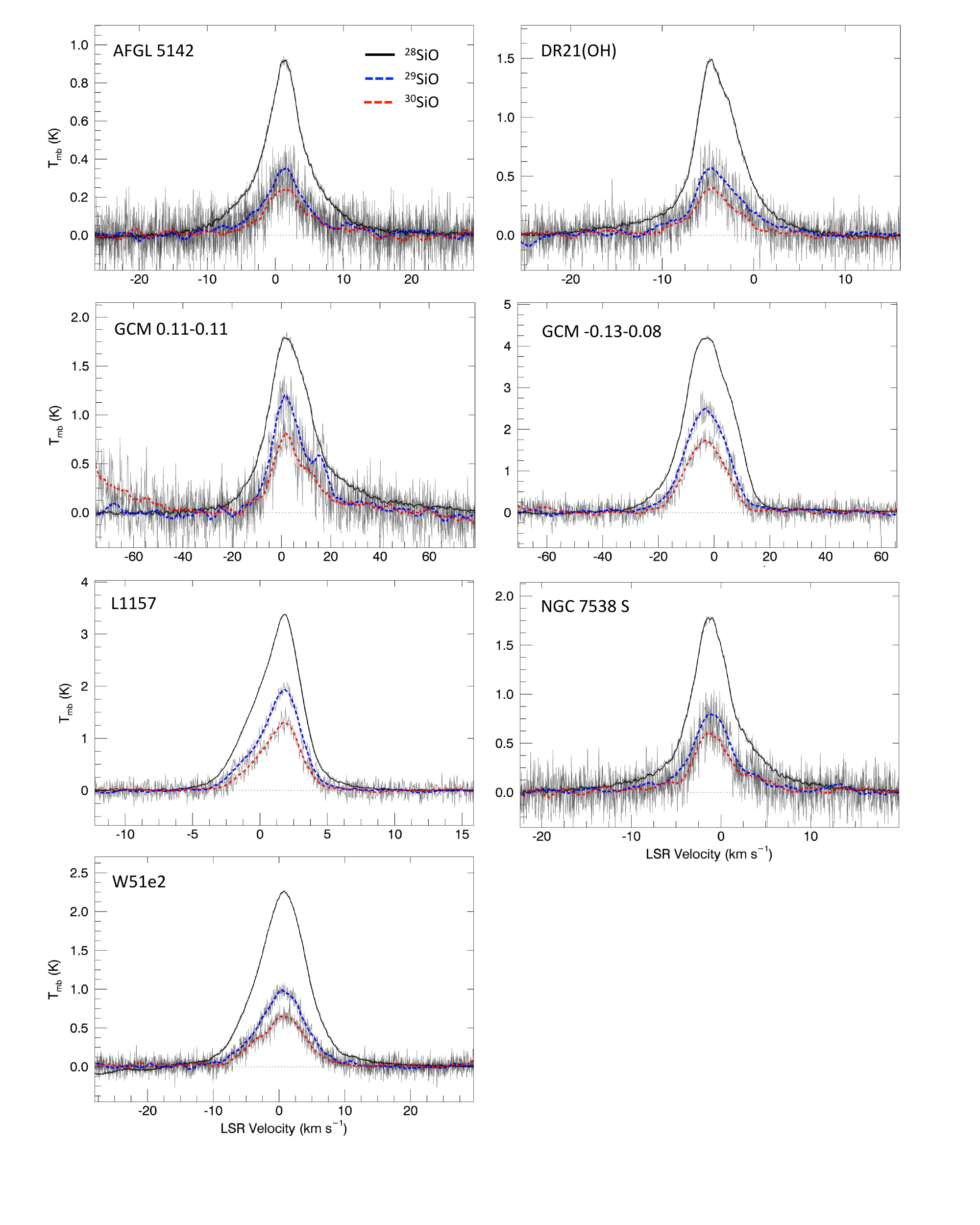}
  \end{center}
  \vspace{-10pt}
  \caption{Flux-calibrated, baseline-corrected $^{28}$SiO, $^{29}$SiO, and $^{30}$SiO emission lines for the seven sources in this study. The grey lines show the unsmoothed, full-resolution spectra while the solid black, dashed blue, and dashed red lines are the smoothed data for  $^{28}$SiO, $^{29}$SiO, and $^{30}$SiO respectively.  Main beam temperatures apply to $^{28}$SiO while the $^{29}$SiO and $^{30}$SiO lines are scaled by a factor of 7 for presentation.  The baseline at the low-velocity extreme of the $^{30}$SiO spectrum for GCM0.11-0.11 is outside the range used to determine the baseline underlying the $^{30}$SiO line itself.}
  \label{fig:spectra}
  \vspace{-5pt}
\end{figure*}

Flux-calibrated spectra with baselines subtracted are shown for each of the seven sources in this study in Figure \ref{fig:spectra}.  The $^{29}$SiO and $^{30}$SiO line intensities are exaggerated by a factor of 7 for presentation.  

\subsection{Interfering Lines}

Extraneous emission lines are seen in most sources, however these extraneous lines generally do not interfere with the SiO lines. Notable exceptions include the six blended 2(0,2) $\to$ 1(0,1)  hyperfine lines of formamide (42385.06 MHz to 42386.68 MHz), which were seen in the $^{30}$SiO spectra of both Galactic center sources. The brightness of the formamide line exceeded that of $^{30}$SiO in both cases and its effects on the $^{30}$SiO lines were removed using the methods described above for baselines. Formamide emission was seen in W51e2 as well, but was rather weak in this source. There was an additional interfering line in W51e2 which appears on the low velocity wing of the $^{30}$SiO line and had to be removed. The poor SNR of the line  made identification difficult, although the line is fairly broad and is possibly a blend of the 13( 3,11) $\to$ 12( 4, 8) EA and 13( 3,11) $\to$ 12( 4, 8) AE emission lines of dimethyl ether at 42371.58 MHz and 42372.16 MHz, respectively.

\begin{figure*} [t]
  \begin{center}
    \includegraphics[width=0.80\textwidth]{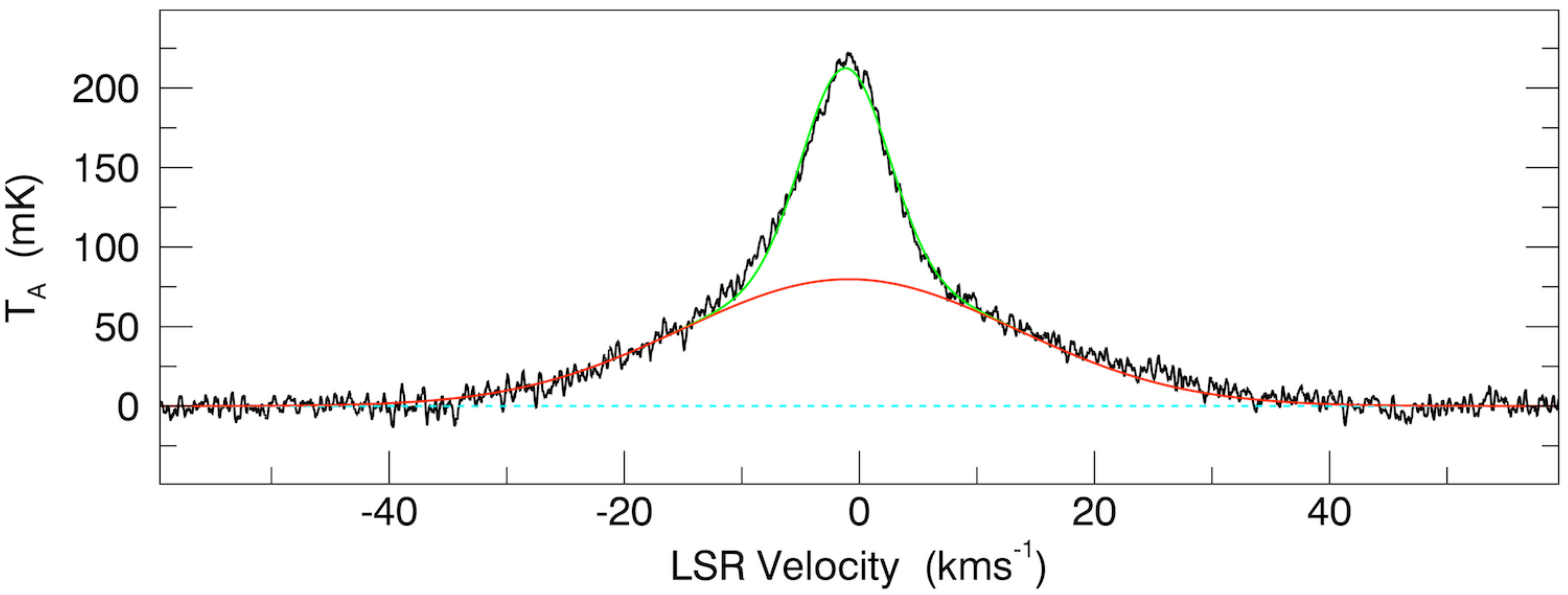}
  \end{center}
  \vspace{-10pt}
  \caption{The $^{29}$SiO emission line from W51e2 with underlying H(83)$\delta$ recombination line fitted (smooth curves) for subtraction. Line intensities are shown in antenna temperature in this figure.}
  \label{fig:rrl1}
  \vspace{0pt}
\end{figure*}

W51e2 also exhibits the H(83)$\delta$ recombination line in the $^{29}$SiO spectrum. The H(83)$\delta$ recombination line lies well within 1 MHz of the $^{29}$SiO emission line, is thermally broadened,  and is easily mistaken as being part of the $^{29}$SiO emission line wings (Figure \ref{fig:rrl1}).    Without removal of this overlapping line the measured [$^{29}$SiO]/[$^{28}$SiO] would be in error by over $40\%$. The H(83)$\delta$ recombination line was effectively removed by using the 'spare' IF to observe the nearby and stronger H(53)$\alpha$ recombination line, which was then used as a template profile to fit and subtract the H(83)$\delta$ line from the $^{29}$SiO spectrum (e.g., Figure \ref{fig:rrl1}). As a precaution, the H(53)$\alpha$  line was monitored in all other sources, although it was only observed in W51e2.  

\subsection{Extracting Column Densities from Line Intensities}
In order to extract isotopologue ratios from line intensities, we forgo the Rayleigh-Jeans approximation (see the Appendix) and express the upper level population column density ratio of secondary (i.e., rare in our application) and primary (abundant in our application)  isotopologues $N_{u}^{s}/N_{u}^{p}$  as
\begin{equation} \label{eq:cd4}
\frac{N_{u}^{s}}{N_{u}^{p}} = \frac{W_{s} \Lambda_{s} {\nu_{u \ell}^{p}}^{3}} {W_{p} \Lambda_{p} {\nu_{u \ell}^{s}}^{3}} \left[ \frac{1 - n_{\gamma}{(T_{\rm crf})} / n_{\gamma}{(T_{\rm ex})_p}}{1 - n_{\gamma}{(T_{\rm crf})} / n_{\gamma}{(T_{\rm ex})_s}} \right],
\end{equation}
where $W_p$ and $W_s$ are the integrated line intensities for the primary and secondary silicon isotopologues, $p$ and $s$ respectively, $ n_{\gamma}(T_{\rm ex})_p$ and $ n_{\gamma}(T_{\rm ex})_s$ are the equivalent photon occupation numbers for the excitation temperatures for isotopologues $p$ and $s$, and $n_{\gamma}(T_{\rm crf})$ is the equivalent photon occupation number for the local continuum radiation  field.  $\Lambda_p$ and $\Lambda_s$ values correct for optical depths for the two isotopologues and are described in the next section. By including the bracketed term we allow for isotopologue-specific subthermal excitation effects.  

Although the flux contribution from the local continuum radiation field, expressed here as $ n_{\gamma}(T_{\rm crf})$, is effectively invariant between isotopologues, feedbacks in the line radiation field, or line trapping, will have a differential effect on emission any time local thermodynamic equilibrium (LTE) does not obtain.  Large dipole moments, and thus large Einstein $A_{ul}$ coefficients for spontaneous emission, raise the probability of line trapping.  Line trapping has the effect of increasing the excitation temperatures of affected transitions relative to the excitation temperature that would occur if the only radiative contribution was the continuum radiation field.   Therefore, for cases where $T_{\rm ex} < T_{\rm kin}$ (where $T_{\rm kin}$ is the kinetic temperature) and where $T_{\rm crf}$ is low (e.g., approaching the CMB),  line trapping can pump up the isotopologue-dependent excitation temperatures and produce inaccuracies  in the derived isotopologue ratios if not accounted for.   

The conditions that foster the isotopologue-selective effects of line trapping can be illustrated using an expression for excitation temperature for a two-level system \citep[e.g.][]{Goldsmith1972}:

\begin{equation} \label{eq:tex}
T_{{\text{\rm ex}}} = \frac{{({T_{{\text{\rm crf}}}} + {T_{{\text{\rm line}}}}) + \frac{h\nu }{{{k_{\text{\rm b}}}}}\frac{{{C_{u \ell}}}}{{{A_{u \ell}}}}}}
{{1 + \frac{{h\nu }}{{{k_{\text{\rm b}}}{T_{{\text{kin}}}}}}\frac{{{C_{u \ell}}}}{{{A_{u \ell}}}}}},
\end{equation}
where $A_{ul}$ is the Einstein coefficient for spontaneous emission and $C_{ul}$ is the collisional de-excitation rate, $T_{\rm line}$ is the radiative contribution to the excitation temperature in the transition, $T_{\rm crf}$ is again the equivalent blackbody temperature of the continuum radiation field, $T_{\rm kin}$ is the kinetic temperature, and the other symbols have their usual meanings.  The collisional de-excitation rate depends on the number density of molecules.  As number density tends to zero, and thus $C_{u \ell}\rightarrow 0$, $T_{\rm ex} \rightarrow (T_{\rm crf} + T_{\rm line})$ (Equation \ref{eq:tex}).  In this case, emission is subthermal and  $T_{\rm line}$  competes with $T_{\rm crf}$ for dominance in determining the excitation temperature $T_{\rm ex}$.  The line temperature is enhanced by line trapping that in turn rises with the abundance, and thus column density, of the emitting isotopologue.  The isotopologue-specific effects are diminished at higher continuum temperatures because of the diluting effects of the isotopologue-independant $T_{\rm crf}$.  Conversely, as the number density tends to infinity, and thus $C_{u \ell}\rightarrow \infty$,  $T_{\rm ex} \rightarrow T_{\rm kin}$ (Equation \ref{eq:tex}), and the system is in LTE.  In this case, there are no isotopologue-specific emission effects due to line trapping.  In summary, Equation (\ref{eq:tex}) shows  that low number densities and low continuum radiation temperatures facilitate isotopologue-specific enhancements in emission due to subthermal excitation and line trapping.  Because the rotational states of SiO are subthermally populated in at least some of our sources \citep[e.g., ][]{Nisini:2007, Amo-Baladron:2009}, and probably in all \citep{Harju:1998}, we evaluated the potential biases in our derived isotopologue ratios attributable to this phenomenon.

\begin{figure*} 
  \begin{center}
    \includegraphics[width=0.8\textwidth]{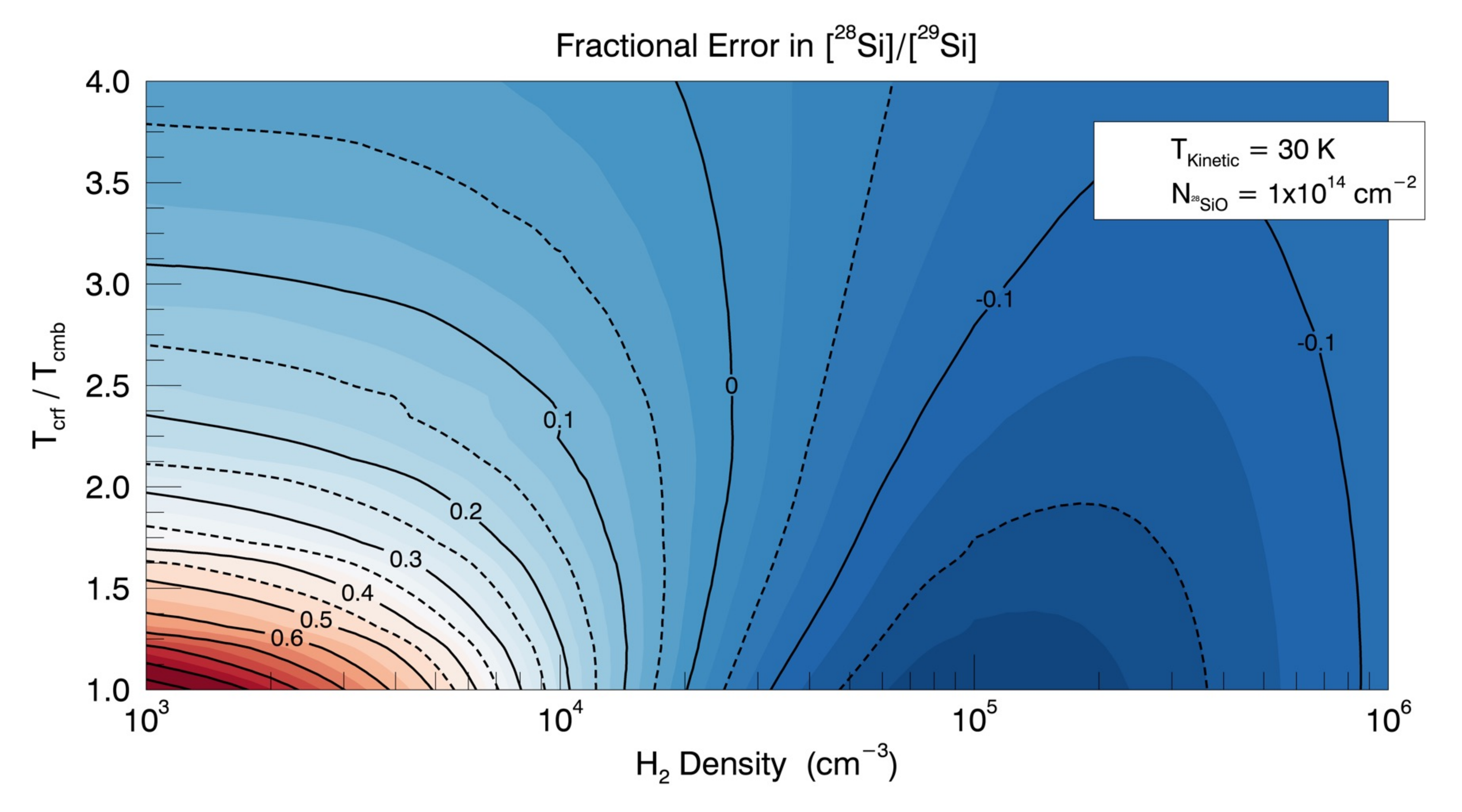}
  \end{center}
  \vspace{-10pt}
  \caption{Contours of errors in SiO isotopologue ratios obtained from integrated J $= 1 \to 0$ emission line areas as a function of collision partner number density and the temperature of the incident continuum radiation field, $T_{\rm crf}$, expressed in multiples of the cosmic microwave background temperature, $T_{\rm cmb}$.  As indicated in the inset, the kinetic temperature of the gas and the column density of $^{28}$SiO are fixed at 30 K and $1\times10^{14}$ respectively. The ratios are fully corrected for optical depth, thus the error is due purely to disparate excitation between the $^{28}$SiO and $^{29}$SiO isotopologues. Solid contours are fractional differences between the $[^{28}$Si] / [$^{29}$Si] extracted from the model and the input parameters (solar in all cases) in increments of $\pm 0.1$. Dashed contours represent midpoints between solid contours, and are plotted only where the magnitude of the fractional error is $< 0.1$, as the gradient in the data is comparatively shallow in that region. Radiative transfer calculations were made with RADEX using the large velocity gradient approximation (van der Tak et al. 2007).}
  \label{fig:radex_plot}
  \vspace{-5pt}
\end{figure*}

RADEX \citep{vanderTak:2013} was used to constrain the magnitude of error induced by divergent excitation temperatures among isotopologues as a function of H$_2$ density and continuum radiation field intensity.  We used the large velocity gradient approximation for the calculations presented here.  Calculations based on the plane-parallel and spherical geometries do not yield appreciably different results from those shown here.  
 Figure \ref{fig:radex_plot} shows contours of fractional deviations in measured optical-depth-corrected isotopologue ratios from the true ratios as a function of collision partner number density and the temperature of the local continuum radiation field.  The kinetic temperature is assumed to be $30$K, but the results are insensitive to the kinetic temperature as long as $T_{\rm kin} \gtrsim 10 {\rm K}$. The contours illustrate that errors well in excess of $20\%$ are expected in low H$_{2}$ density, low continuum flux environments (e.g., for $\rm H_2$ number densities $n_{\rm H_2} <  5\times 10^3$ and $T_{\rm crf}$ less than twice the CMB) if the excitation effects go unrecognized.  Published descriptions of the targets in our study report strong sources of millimeter continuum emission in proximity to the SiO emission sources, typically in the form of either ultra-compact H~II regions or winds from nearby high-mass young stellar objects \citep[e.g., ][]{Luisi:2016, Araya:2009, Hunter:1999, Zapata:2009}.  Therefore, the temperatures of the continuum radiation within our sources are by all evidence well in excess of the CMB, mitigating isotope-specific excitation effects.  Similarly, for the SiO sources reported here $10^4 \lesssim n_{\rm H_2} \lesssim 10^6$ $\rm cm^{-3}$ and so the environs of these sources  correspond to conditions where systematic errors are likely to be $< 10\%$ (Figure \ref{fig:radex_plot}), commensurate with the measurement errors.  While radiation field effects are an important consideration, they do not appear to be sufficient to significantly alter the isotopologue ratios extracted from the data in this study.

\subsection{Optical Depth Corrections}

Historically, SiO emission has been assumed to be optically thin (Wolff, 1980) due to the modest brightness of the observed lines.  However, Penzias (1981) was quick to demonstrate that SiO thermal emission often contravenes this assumption, and the same was found to be true for this survey. Many studies of interstellar isotope ratios categorize emission lines into one of two groups: optically thin where optical depth at line center ($\tau_o$) is much less than unity or optically thick where $\tau_o \gg 1$ \citep[e.g., ][]{Adande:2012,Milam:2005,Savage:2002}. Lines are then analyzed in the appropriate limit. This approach has the convenience of simplicity and is a concession to the difficulty in assessing optical depth in radio emission lines in general \citep{Goldsmith:1999}. 

Many emission lines, however, will not be patently either thick ($\tau_o \gg 1$) \emph{or} thin ($\tau_o \ll 1$), and instead are likely to exhibit some finite intermediate values for $\tau$ \citep[e.g., ][]{Savage:2002,Milam:2005,Penzias:1981}.  This should be especially true for emission from dense gas tracers like those in this study, where even moderately bright lines from highly subthermal populations have appreciable optical depths.  The limits of thin or thick will therefore result in significant errors. Use of the thin limit appears particularly problematic as error grows rapidly with optical depth, reaching $\sim 10\%$ for even a moderate $\tau$ at line center of $0.2$. 

\begin{figure*}[t] 
  \begin{center}
    \includegraphics[width=0.9\textwidth]{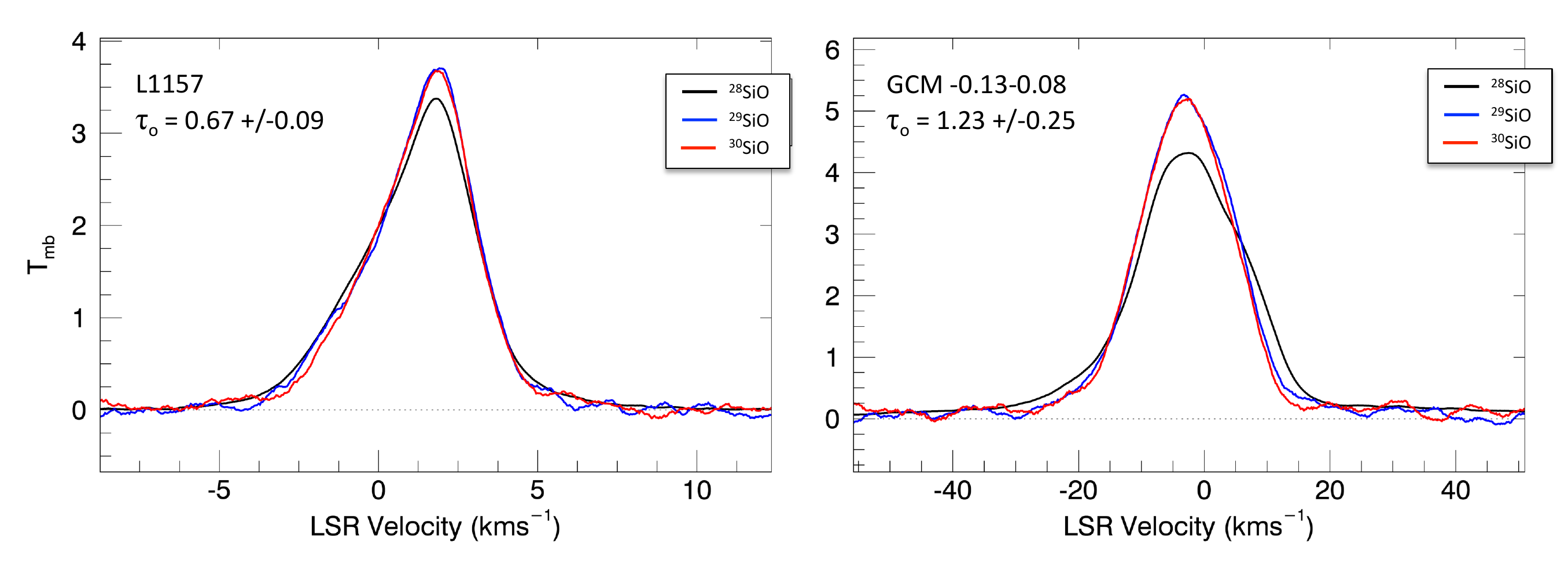}
  \end{center}
  \caption{Area-scaled emission line profiles for the $v = 0, J = 1 \to 0$ transitions for the three SiO silicon isotopologues observed in L1157 (left) and GCM-0.13-0.08 (right). The black lines are the $^{28}$SiO lines. The red and blue profiles are the $^{29}$SiO and $^{30}$SiO lines, respectively.  Each line has been scaled by integrated intensity relative to the$^{28}$SiO integrated intensity as in Equation (\ref{eq:fix1}). The disparities between line-center $T_{\rm mb}$ values for the $^{28}$SiO lines and the scaled $^{29}$SiO and $^{30}$SiO lines are indicative of appreciable optical depths in the $^{28}$SiO lines (see text).} 
  \label{fig:sf1}
\end{figure*}

We developed a method for estimating optical depth for the major SiO isotopologue in this study based on comparisons of line shapes.  The underlying foundational premise is that high optical depths manifest as broadening in the $\rm ^{28}SiO$ line relative to the rarer isotopologue lines (i.e., we assume in Equation \ref{eq:cd4} that $\Lambda_{29} = \Lambda_{30} = 1$ but allow for $\Lambda_{28} \ge 1$) that is obvious when the emission lines for the different isotopologues are scaled by area.  With this method, optical depths in the $^{28}$SiO emission lines are determined by analyzing differences between the $^{28}$SiO and  $^{29}$SiO and/or $^{30}$SiO lineshapes for the same source under the assumption that the latter is effectively optically thin. First, the FWHM breadth of the $^{28}$SiO emission line is determined by fitting a Voigt profile to the line in main beam temperature - LSR velocity ($v_r$) space.   All three isotopologue lines are then integrated over $\Delta v_r = \pm 3$ FWHM (as defined by the $\rm ^{28}SiO$ line) from line center. The lines of the two rarer isotopologues, $^{29}$SiO and $^{30}$SiO,  are scaled by the ratio of the abundant-to-rare integrated line areas.  For example the scaled main beam temperatures for the $\rm ^{29}SiO$ lines are  
\begin{equation} \label{eq:fix1}
T^{\rm \ Scaled}_{\rm mb, ^{29}SiO }(v_r) = T_{\rm mb, ^{29}SiO }(v_r) \left( \frac{\int T_{\rm mb, ^{28}SiO }(v_r)  \, {d}v_r}{\int T_{\rm mb, ^{29}SiO }(v_r)  \, {d}v_r} \right).
 \end{equation}
The scaled $^{29}$SiO and $^{30}$SiO lines are superimposed on the $^{28}$SiO line (e.g., Figure \ref{fig:sf1}). Because the scaled $^{29}$SiO and $^{30}$SiO lines trace one another within uncertainties (they have comparable, presumably low, optical depths based on their normal abundances), any broadening in the$^{28}$SiO line is immediately obvious as an apparent deficit in main beam temperature at line center (Figure \ref{fig:sf1}).  The ratio of the $^{i}$SiO main beam temperatures at line center to the integrated line intensities, $\Gamma_i$:
\begin{equation} \label{eq:fix2}
\Gamma_i = \frac{T^{o}_{\rm mb, ^{i}SiO }}{\int T_{\rm mb, ^{i}SiO }(v_r) \, {d}v_r}
 \end{equation}
we refer to as the ``shape parameter''.  We use this shape parameter to quantify optical depths.  Both intensity at line center and integrated area of a spectral line are non-linearly dependent upon optical depth, with peak intensity at line center exhibiting a stronger dependence than area.  This is because optical depth varies across the line profile with the line wings remaining relatively thin even as $\tau$ at line center increases. As optical depth increases, the shape parameter $\Gamma_i$ decreases (the profile shape becomes fatter).  The optical depth of an emission line can therefore be determined by comparing the line shape parameter of the suspected optically thick line (for the abundant isotopologue) with that for a line that is presumed to be optically thin (corresponding to the rare isotopologues).  For moderate optical depths we find that the optical depth at line-center for the optically thick line is linearly proportional to the fractional difference between shape parameters for the thick and thin lines:

\begin{equation} \label{eq:fix3}
\tau_o \propto   \frac{\Gamma_{\rm thin}}{\Gamma_{\rm thick}}-1.
\end{equation}
 Evaluation of synthetic data indicates that an empirically-derived proportionality constant value of $5$ in Equation (\ref{eq:fix3}) produces accuracy in derived $\tau_o$ values within $\sim 5\%$ when $\tau_o$ is near $\sim 2$ and within $\sim 2\%$ when $\tau_o$ is near $1$.  All of the line-center optical depths obtained as part of this study are $< 1.5$.   We tested this process using synthetic lines and find the accuracy to be robust against irregular line profiles and even cryptically overlapping velocity components from separate clumps within a complex source.  It is worth emphasizing that influences of velocity structure on line shapes are not isotope specific, and our forward calculations verify that optical depth effects alone result in the departures from line shape coincidence when normalized to area.  A caveat is that there are hypothetical circumstances where one can imagine localized velocity features that affect the rare isotopologues differently than the abundant species, but these will be pathological circumstances.  
 
Another caveat is that, if there are strong gradients in excitation temperature along the line of sight, then  an optically thick line for the abundant species will favor the foreground values of excitation temperature for that species only, leading to an error in the abundance ratio.  This is a known and important effect for very optically thick lines like those of $\rm ^{12}C^{16}O$; indeed, in the absence of a velocity gradient along the line of sight, one is typically observing only the surface layers of a cloud in the most abundant isotopologue of CO.  However, for SiO, this effect is minimized because of the comparatively modest values of optical depth for the SiO lines.  Furthermore, there is little reason to expect strong line-of-sight excitation gradients in the kinds of sources that give rise to SiO emission; the SiO molecules are likely intermixed with the shocks that liberate or form them. 
 
 Our observation is that failures of Equation (\ref{eq:fix3}) require models that invoke rather unlikely circumstances. Our forward calculations demonstrate further that details of line shapes (e.g., skewness) do not significantly alter the relationship between $\Gamma_i$ and line-center optical depth as long as the line profile is not flat-topped.
  
The optical-depth correction for measured column densities for $^{28}$SiO takes the form

 \begin{equation} \label{eq:fix4}
 \Lambda_{28} = \frac{N^{28}_{\rm corrected}}{N^{28}_{\rm uncorrected}} = \frac{\int \tau_{v_r} dv_r}{\int \left(1-\exp(-\tau_{v_r})\right)\, dv_r}.
\end{equation}
The integrals in Equation (\ref{eq:fix4}) are obtained from the optical depths at line center from Equation (\ref{eq:fix3}) and the line profile functions defined by the $^{29}$SiO lines (assumed to be optically thin).  Although we derived this correction factor independently (see Appendix), one could use equations 83-85 from  \cite{Mangum2015} to derive the same expression,  although they instead appeal to an expression for $\Lambda_i$ attributable to  \cite{Goldsmith:1999} in their text that is not correct for application with Equation (\ref{eq:cd4}).  

We use Equations (\ref{eq:fix2}), (\ref{eq:fix3}), and (\ref{eq:fix4}) to determine optical depths for the $^{28}$SiO lines for all sources reported here. In all cases the two derived $\tau_o$ values based on the $^{29}$SiO and $^{30}$SiO shape parameters are in agreement within uncertainties; we used the SNR-weighted average of the two for the $\tau_o$ value reported for each source.  Values for $\tau_o$ differ for the different sources, with values ranging from below detection to  slightly greater than unity.  The $^{28}$SiO lines from DR21(OH) and AFGL 5142 have optical depths below detection, with a noise-limited detection limit of $\tau_o \approx 0.2$. The $T_{\rm mb}$ values for the $^{28}$SiO lines are less than 1K in both sources.  The peak $T_{\rm mb}$ values for the $^{28}$SiO emission line in W51e2 is $\sim$ 3K and is also relatively optically thin, with an estimated optical depth of $\tau_o \approx 0.4$. The two Galactic center sources and L 1157, by contrast,  all show evidence for appreciable optical depths in the main $^{28}$SiO emission line, with $\tau_o$ values of 1.0, 1.2, and 0.7, respectively (Table 2). 

\subsection{Evaluation of Uncertainties}

In order to account for both measurement uncertainties and the uncertainties imparted by the estimates of optical depth, the entire data reduction pipeline and correction scheme for each source was subjected to a Monte Carlo error analysis.  For this analysis, random draws were made from each channel comprising a spectrum.  The RMS values defined by the off-peak data were used to define the standard deviations about the measured values.  The measured values were taken as the means for the random draws in order to preserve line shapes.  The use of the measured values as means (rather than smoothed values) results in an over estimation of uncertainties in the derived isotopologue ratios of $\sim 3$ to $\sim 5$\%.  The result is two hundred thousand instances of each SiO line for each source.   These lines are used as the input for the data reduction, including the estimates of optical depth.  The limits for defining baselines were also varied for each random draw though we find that the details of the baseline selection yield negligible contributions to the overall uncertainties.  

The corrections for optical depth in the $^{28}$SiO lines (Table 2) generally increase the uncertainties in isotopologue ratios by factors of approximately 2 to 3.  Because of the additional uncertainty in the abundant isotologue column densities, the correlation coefficients between the $[^{29}{\rm SiO}]/[^{28}{\rm SiO}]$ and $^{30}{\rm SiO}]/[^{28}{\rm SiO}]$ ratios increase from $<0.1$ to $ 0.85 \pm 0.2$ in all of the sources.

\section{Results}

\begin{figure}
\begin{center}
    \includegraphics[width=0.45\textwidth]{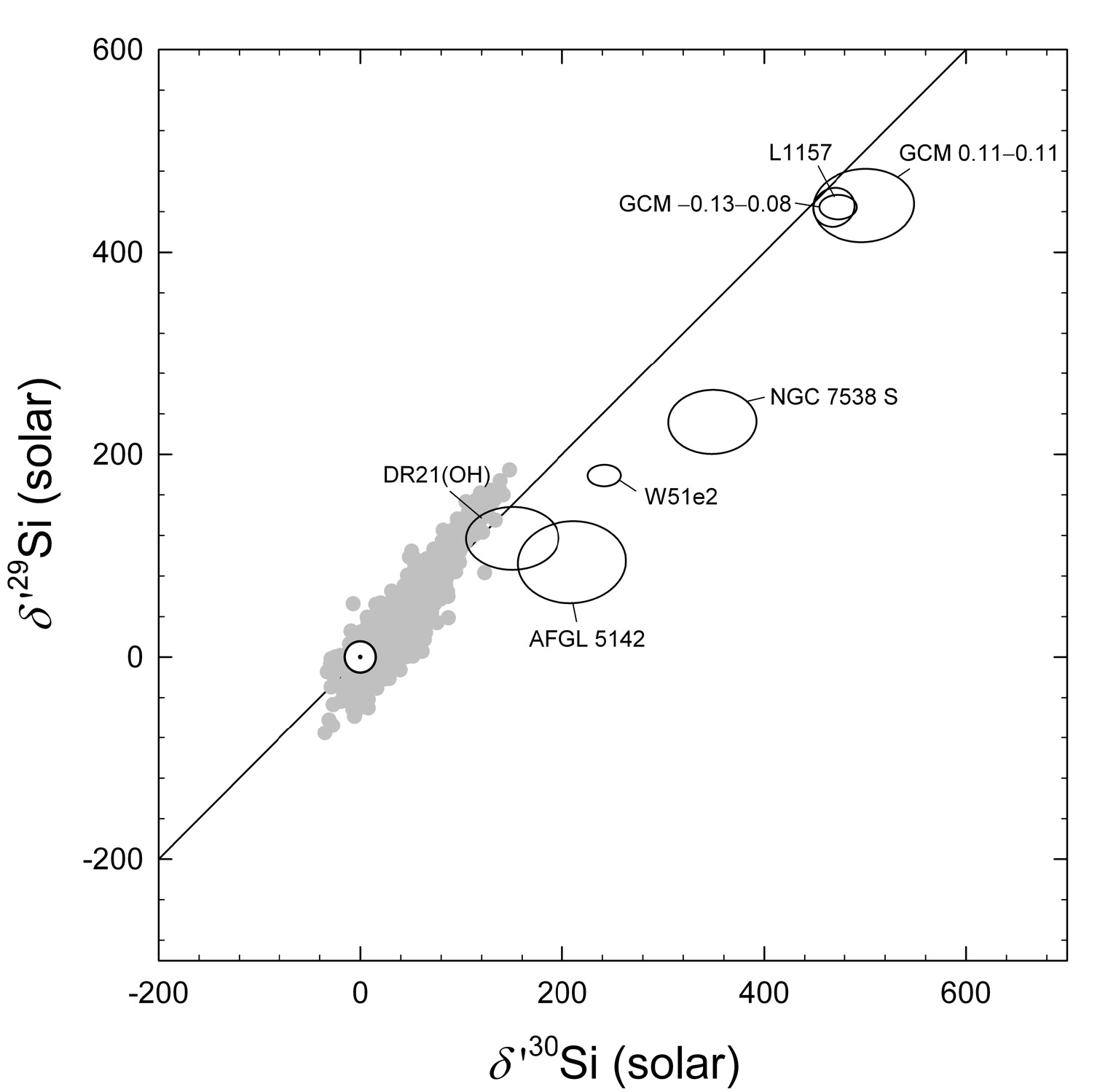}
  \end{center}
  \vspace{-10pt}
  \caption{Uncorrected SiO silicon isotope abundance ratios for the seven sources observed as part of this survey. Mainstream SiC grain data are shown for reference (grey circles). The solid line is the slope-unity line through the solar composition. The white circle with dot indicates present-day solar abundances and defines the origin. Error ellipses are 1$\sigma$.}
  \label{fig:data1} 
  \vspace{-5pt}
\end{figure}

\begin{figure} 
\begin{center}
    \includegraphics[width=0.45\textwidth]{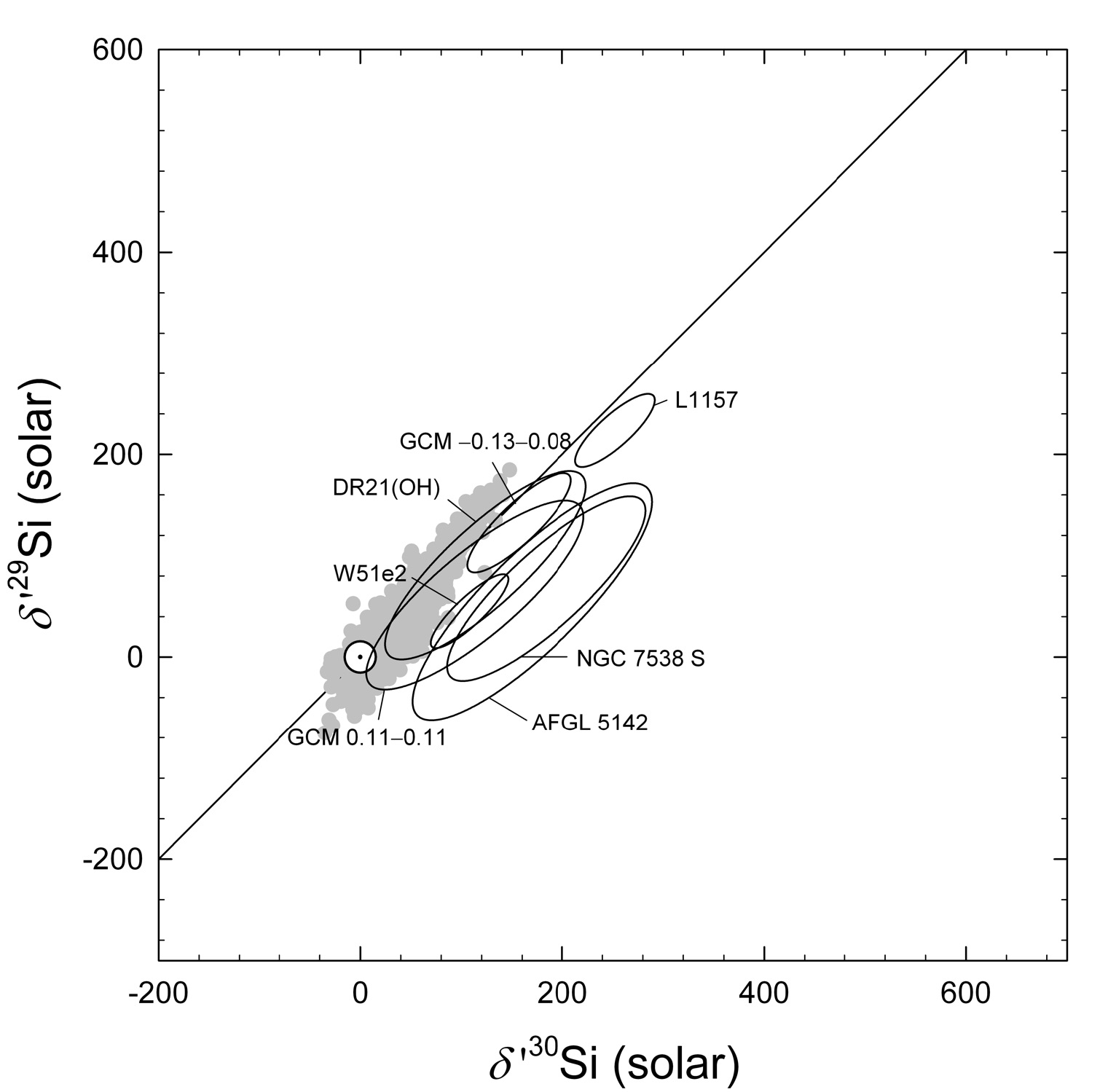}
  \end{center}
  \vspace{-10pt}
  \caption{SiO silicon isotope ratios after correcting for optical depth effects. Error ellipses are 1$\sigma$ determined by Monte Carlo simulations including the uncertainty in the optical depth corrections.}
  \label{fig:data2}  
  \vspace{-5pt} 
\end{figure}

A summary of the results is given in Table 2 and shown in Figures \ref{fig:data1} and \ref{fig:data2}. Our uncorrected data exhibit a spread up and down the slope-1 line in Si three-isotope space, anchored by the two Galactic center sources  and crudely  resembling the predictions from GCE (Figure \ref{fig:data1}).  The trend with $R_{\rm GC}$ is broken by the high [$^{29}$SiO]/[$^{28}$SiO] and [$^{30}$SiO]/[$^{28}$SiO] ratios for L1157 at solar $R_{\mathrm{GC}}$.  However, correcting for optical depth removes the spread in data, resulting instead in a clustering of the data spanning the range defined by the mainstream SiC presolar grain trend (Figure \ref{fig:data2}).  We find, not surprisingly, that optical depths on the order of unity can strongly bias extracted isotope ratios. These results indicate that uncorrected effects of opacities were  responsible for the prior evidence for high [$^{29}$SiO]/[$^{28}$SiO] and [$^{30}$SiO]/[$^{28}$SiO] ratios in the present-day ISM relative to solar and meteoritical values \citep{Penzias:1981,Wolff:1980}.  The prior measurements were suggestive of GCE over the $\ge 4.6$ Gyrs since the birth of the Sun and the formation of the presolar SiC grains. Our new results suggest instead that  silicon isotope ratios have been minimally affected by GCE over this time interval.   

\begin{deluxetable*}{lccccccccc}
\tabletypesize{\footnotesize} 
\tablecolumns{10} 
\tablecaption{Corrected and uncorrected SiO isotopologue ratios.}
\tablehead{
\colhead{}  &  \colhead{} &  \colhead{}  &  \multicolumn{2}{c}{Uncorrected Ratios}  &  \colhead{}  &  \multicolumn{2}{c}{Corrected Ratios}  &   \multicolumn{2}{c}{Relative to Solar}   \\ 
\cline{4-10}  \\
\colhead{Source}  &  \colhead{$\tau_0$}  &  \colhead{$\Lambda_{28}$}  &  \colhead{${[^{28}\mathrm{Si}]}/{[^{29}\mathrm{Si}]}$}  &  
\colhead{${[^{28}\mathrm{Si}]}/{[^{30}\mathrm{Si}]}$}  &  \colhead{}  &  \colhead{${[^{28}\mathrm{Si}]}/{[^{29}\mathrm{Si}]}$}  &  
\colhead{${[^{28}\mathrm{Si}]}/{[^{30}\mathrm{Si}]}$}  &  \colhead{$\delta^{\prime 29}\mathrm{Si}$} & \colhead{$\delta^{\prime 30}\mathrm{Si}$} }
%% All data must appear between the \startdata and \enddata commands
\startdata
DR21 (OH)         &  0.08 $\pm$ 0.21  &  1.03 $\pm$ 0.08  &  17.53 $\pm$ 0.54  &  25.73 $\pm$ 1.18  &   &  18.06 $\pm$ 1.68  &  26.52 $\pm$ 2.65  &  90 $\pm$ 93  &  124 $\pm$ 99 \\
L1157 B1            &  0.67 $\pm$ 0.09  &  1.25 $\pm$ 0.04  &  12.63 $\pm$ 0.15  &  18.61 $\pm$ 0.35  &   &  15.76 $\pm$ 0.57  &  23.23 $\pm$ 0.92  &  223 $\pm$ 36  &  252 $\pm$ 39 \\
NGC 7538 S       &  0.49 $\pm$ 0.22  &  1.18 $\pm$ 0.08  &  15.62 $\pm$ 0.49  &  21.10 $\pm$ 0.92  &   &  18.48 $\pm$ 1.69  &  24.97 $\pm$ 2.47  &  67 $\pm$ 91  &  184 $\pm$ 98 \\
AFGL 5142         &  0.12 $\pm$ 0.26  &  1.04 $\pm$ 0.09  &  17.95 $\pm$ 0.73  &  24.26 $\pm$ 1.32  &   &  18.77 $\pm$ 2.22  &  25.37 $\pm$ 3.03  &  55 $\pm$ 117  &  170 $\pm$ 119 \\
W51e2                &  0.39 $\pm$ 0.09  &  1.14 $\pm$ 0.04  &  16.47 $\pm$ 0.17  &  23.46 $\pm$ 0.39  &   &  18.83 $\pm$ 0.69  &  26.84 $\pm$ 1.03  &  45 $\pm$ 36  &  108 $\pm$ 38 \\
GCM0.11-0.11         &  1.23 $\pm$ 0.25  &  1.47 $\pm$ 0.10  &  12.61 $\pm$ 0.46  &  18.17 $\pm$ 0.91  &   &  18.61 $\pm$ 1.74  &  26.82 $\pm$ 2.90  &  61 $\pm$ 93  &  113 $\pm$ 108 \\
GCM -0.13-0.08  &  0.97 $\pm$ 0.13  &  1.37 $\pm$ 0.05  &  12.63 $\pm$ 0.24  &  18.69 $\pm$ 0.39  &   &  17.27 $\pm$ 0.85  &  25.56 $\pm$ 1.31  &  133 $\pm$ 49  &  157 $\pm$ 51 \\
\enddata
\end{deluxetable*}

\begin{figure} 
\begin{center}
    \includegraphics[width=0.45\textwidth]{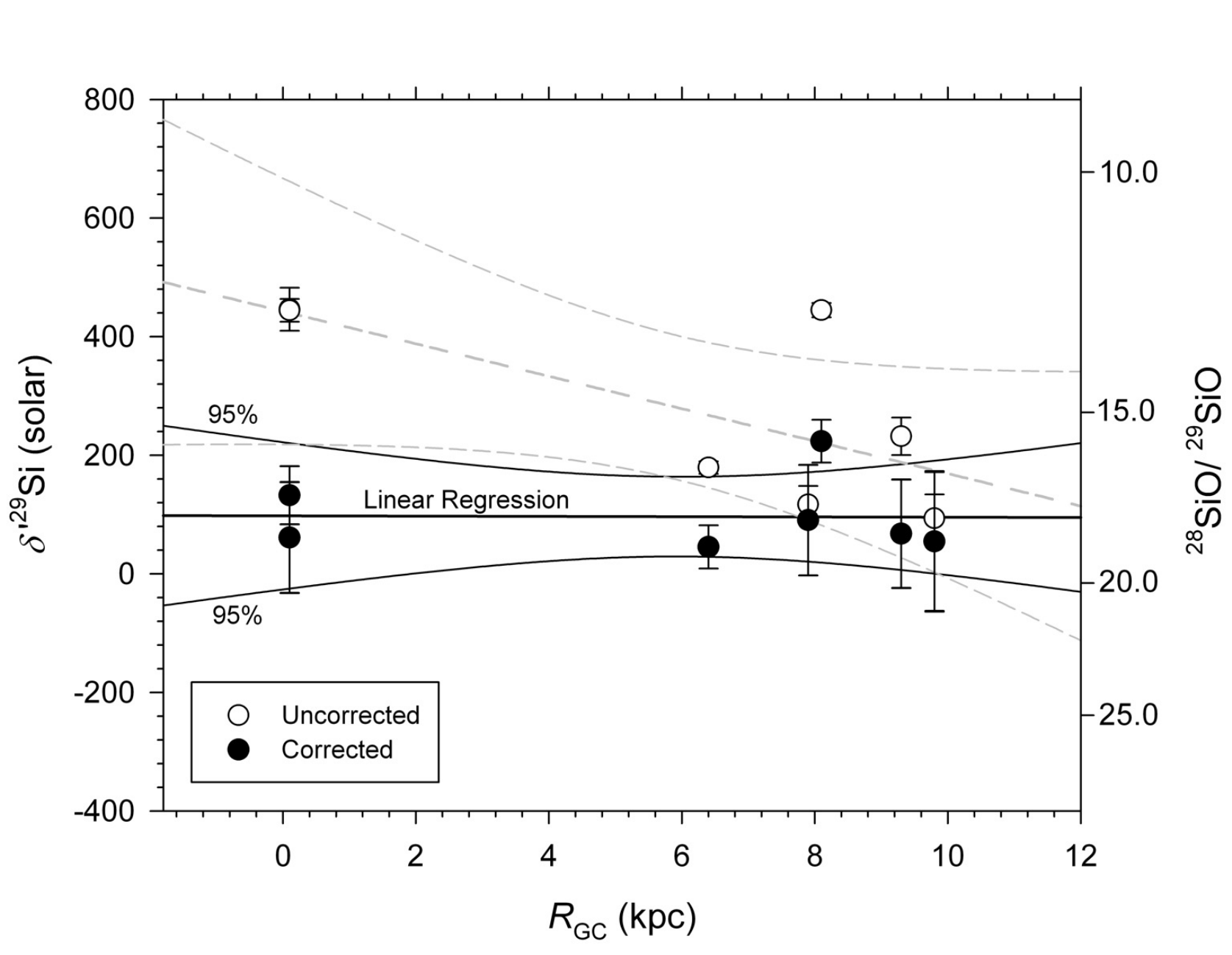}
  \end{center}
  \vspace{-10pt}
  \caption{
  [$^{29}$SiO]/[$^{28}$SiO] in permil vs. Galactocentric distance from this study. Uncorrected data are shown as open symbols.  Data corrected for optical depth are shown as black symbols.  Error bars are 1$\sigma$.  Linear regression for the uncorrected data (grey) and corrected data (black) are shown together with 95\% confidence bands.  The corresponding [$^{28}$SiO]/[$^{29}$SiO] ratios are shown on the right-hand ordinate.
  }
  \label{fig:data3}  
  \vspace{-5pt} 
\end{figure}

Correcting for optical depths removes the evidence for a variation in silicon isotope ratios with  $R_{\rm GC}$  (Figure \ref{fig:data3}).  Regression of the uncorrected $\delta ^{\prime 29}{\rm Si}$ values vs $R_{\rm GC}$ gives a negative slope (slope $= -27 \pm 12$ per mil kpc$^{-1}$, Figure \ref{fig:data3}) while regression of the corrected data yields a slope indistinguishable from zero (slope $= -0.2 \pm 6.8$ per mil kpc$^{-1}$, Figure \ref{fig:data3}  ).  

The mean corrected [$^{28}$SiO]/[$^{29}$SiO] ratio for the sources reported here is $17.9 \pm 1.1~(1\sigma)$ and is $9\%$ lower than the solar value of 19.7 (i.e., the average measured values are enriched in $^{29}$SiO relative to solar by $97$ per mil).  The mean of the SiO measurements is slightly further up the slope-1 line in Figure \ref{fig:data2} than the mean of the presolar SiC grains, although the difference is within $2\sigma$ defined by the spread in SiC data (mean [$^{28}$SiO]/[$^{29}$SiO] $=18.9\pm 0.5\, (1\sigma)$ for SiC grains).  The spread in Si isotope ratios from $R_{\rm GC} =10$ kpc to the Galactic center is comparable to the spread in isotope ratios observed in presolar mainstream SiC grains (Figure \ref{fig:data2}) but considerably smaller than predictions based on the apparent variations in oxygen isotope ratios (Figure \ref{fig:co1}).

\section{Discussion}

\subsection{Secondary/Primary Si Isotope Ratios}

The somewhat higher $[\rm ^{29}Si]/[^{28}Si]$ and $[\rm ^{30}S]/[^{28}Si]$ ratios of the present-day ISM relative to solar values presumably represents GCE over the last 4.6 Gyrs.   The finding that there  is no resolvable variation in silicon isotope ratios across the Galaxy is important because it conflicts with expectations from oxygen and carbon secondary/primary isotope ratio trends.   The implication is that in the present-day Milky Way, stars are forming with similar average silicon isotope ratios regardless of their distance from the Galactic center.  

The explanation for the lack of a radial gradient in this isotope system remains elusive.  One possibility is mixing by radial gas flows \citep{Tinsley:1978}.  Simulations suggest that spiral arm - bar resonances and infall of gas can result in flattening in metallicity gradients with $R_{\rm GC}$ in both stars and gas on timescales of $< 1$ Gyr \citep{Minchev:2011,Cavichia:2014}.  If mixing is the cause of the flat gradient for silicon isotope ratios, it would imply that gradients in metallicity and gradients in other isotopic indicators of GCE have also been at least partially flattened by mixing.  

An alternative explanation is a temporal change in the sources of silicon isotopes that is peculiar to silicon.  \cite{Zinner:2006} reconstructed the GCE of Si isotopes using the measured isotope ratios in presolar SiC grains of type Z and models to filter out the nucleosynthetic effects of the AGB stellar progenitors of this rare class of SiC grains.  They  concluded that there was a rapid rise in secondary/primary Si isotope ratios early in the Galaxy followed by a leveling off in the rate of change in these ratios when total metallicity ($Z$) began to exceed $ 0.01$.  These authors suggested that late additions of nearly pure $\rm ^{28}Si$ by Type Ia supernovae, as suggested by \cite{Gallino:1994}, may have contributed to the slowing in the rise of  $[\rm ^{29}SiO]/[^{28}SiO]$ and $[\rm ^{30}SiO]/[^{28}SiO]$ with metallicity (and time).  In this scenario, the addition of $\rm ^{28}SiO$ to the Galaxy was delayed because of the relatively long timescales required for the evolution of Type Ia supernova progenitors  \citep[e.g.][]{Tsujimoto:1995}. Late addition of $\rm ^{28}SiO$ could have minimized the change in Galactic $[\rm ^{29}Si]/[^{28}Si]$ and $[\rm ^{30}S]/[^{28}Si]$ over time, perhaps explaining the modest difference between the solar value and the average ISM today.  

Suppression of a Si isotope gradient with $R_{\rm GC}$ by a rise in the influence of Type Ia supernovae would require that the relative contribution of $\rm ^{28}Si$ from these products of white-dwarf-bearing binary systems is greater towards the Galactic center, counterbalancing the overall rise in metallicity and secondary isotope formation with decreasing $R_{\rm GC}$. \cite{Scannapieco:2005} developed a model for Type Ia formation rate in terms of star formation rate and total stellar mass, implying an overall increase in the rate of Type Ia formation towards the Galactic center.  An accelerated decrease in [O/Fe] with increasing [Fe/H] toward the Galactic center is a signature of the influence of Type Ia supernovae owing to the large mass of Fe released in Type Ia events \citep[e.g., ][]{Matteucci:2006}. It is conceivable that an analogous excess in Type Ia-produced $\rm ^{28}Si$ may exist progressively towards the Galactic center.

\subsection{Secondary/Secondary Si Isotope Ratios}

The weight of the data for the seven sources is displaced from the presolar mainstream SiC data, with the former having higher $[\rm ^{30}SiO]/[^{28}SiO]$ for the same $[\rm ^{29}SiO]/[^{28}SiO]$ ratios (i.e., the SiO data lie to the right of the SiC data in Figure \ref{fig:data2}). This displacement, representing a higher $[\rm ^{30}SiO]/[^{29}SiO]$ than both solar and the presolar mainstream SiC grains, could reflect a difference in the GCE of the two secondary silicon isotopes. Presolar SiC grains of types Y and Z have large excesses in $[\rm ^{30}SiO]/[^{29}SiO]$ resulting from neutron capture in low-mass, low-metallicity AGB stars \citep{Zinner:2006}.  These grains represent a mechanism for altering the ratio of secondary silicon isotopes with time.  However, the AGB source of Si is thought to be relatively minor \citep{Clayton:2003,Timmes:1996} and so the influence of AGB stars in shifting ISM $[\rm ^{30}SiO]/[^{29}SiO]$ over time is expected to be limited.

Enhancements in $[\rm ^{30}SiO]/[^{29}SiO]$ could be indicative of a mass-dependent isotope partitioning (fractionation) because mass-dependent fractionation trends in Figure \ref{fig:data2} have slopes of approximately $1/2$ rather than unity, altering the secondary/secondary $[\rm ^{30}SiO]/[^{29}SiO]$ ratios; the offset between the presolar SiC data and the ISM data could be explained if the the ISM SiO experienced mass-dependent heavy isotope enrichment.  

SiO is commonly associated with both C-type and J-type shocks in the ISM, where it is produced through non-thermal sputtering processes with heavy neutral species (He, C, O \& Fe), as well as vaporization by grain-grain collisions. Si-bearing species are sputtered from both the cores and mantles of grains, and enter the gas phase as either SiO or neutral Si, depending on the grain composition and shock velocity \citep{Martin:2009, Caselli:1997, Schilke:1997, Ziurys:1989}.  SiO sputtering yields are known to vary with impact energy and are mass dependent; sputtering should result in mass-dependent isotope fractionation in which the heavy isotopes are enriched in the condensed phase residues.  The magnitudes of the isotope fractionations associated with sputtering of silicate grains like olivine, the most likely hosts for Si in the ISM,   are not well constrained in the environments studied here. 

Although grain loss is believed to be non-thermal in the environments observed in this study, there may be parallels in the isotope systematics of thermal evaporation/sublimation and sputtering given that the rate of the latter depends on a mass-dependent cohesive binding energy barrier. Thermal evaporation or sublimation of condensed silicates is known to cause Si isotope enrichment in the evaporative residues up to a few per cent where the distillation is extreme.  These results are well documented from  theory, experiments, and observations of meteoritical materials \citep{Shahar:2007}.    The effects of partial evaporation of grains would leave the gas depleted in the heavy, secondary Si isotopes and the residual grains enriched in the heavy isotopes with the relative changes in  [$^{30}$SiO]/[$^{28}$SiO] ratios being twice those for  $[\rm ^{29}S]/[^{28}Si]$ as a consequence of the different vibrational frequencies of ruptured bonds (vibrational frequencies are proportional to the inverse square root of reduced mass).  For example, evaporation of 90\% of the Si from a typical silicate should yield an increase in $[\rm ^{29}S]/[^{28}Si]$ of $\sim 4\%$ in the residual condensed material and a corresponding increase in $[\rm ^{30}SiO]/[^{28}SiO]$ of $\sim 8\%$ \citep{Richter:2007, Shahar:2007}.  This magnitude of fractionation would be sufficient to explain the offset between the SiC and ISM data.   However, the sign is wrong for a simple single stage of grain evaporation.  Rather than the SiO gas being depleted in the heavy isotopes, our data imply enrichment relative to the older SiC grains (Figure \ref{fig:data2}). If grain evaporation/sublimation is an explanation for the offset between SiC grains and SiO gas in Figure \ref{fig:data2}, it would require extreme distillation by Rayleigh-like processes or multiple discrete steps of partial Si loss so that the SiO we measure derives from grains that had a prior history of evaporation and hence heavy isotope enrichment.   These multiple steps cannot have led to complete grain loss because fractionations are only possible where Si is retained in evaporative residues.  
 
Silicon monoxide can be released into the gas phase directly by thermally-driven sublimation or evaporation of silicate grains \citep{Nichols1995}.  In the case of sputtering \citep{May2000}, SiO can form in the gas by reactions between Si and either molecular oxygen or the hydroxyl radical:
\begin{eqnarray}
\mathrm{Si} + \mathrm{O_2} &\to& \mathrm{SiO} + \mathrm{O} \\
\mathrm{Si} + \mathrm{OH^{\cdot}} &\to& \mathrm{SiO} + \mathrm{H^{\cdot}}.
\end{eqnarray}
The SiO/H$_2$ abundance ratio in shocked regions is enhanced by up to $10^5$ relative to the ambient medium, but quickly declines in the cooling post-shock material.  The rates of these gas-phase reactions depend on collision frequencies, raising the possibility that the product SiO might be affected by mass-dependent fractionation relative to the sputtered silicon as a result of the collision frequencies of the silicon atoms that are proportional to $m^{-1/2}$ where $m$ is the atomic mass.  Here again, the sign of the expected shifts is the opposite of that required to explain the offset between SiC grains and SiO gas in Figure \ref{fig:data2}. 

The archetypal destruction pathway of SiO to form SiO$_2$ is the reaction 
\begin{equation}
\mathrm{SiO} + \mathrm{OH^{\cdot}} \to \mathrm{SiO_2} + \mathrm{H^{\cdot}}
\end{equation}
occurring in the post-shock gas, where OH$^{\cdot}$ is abundant \citep{Schilke:1997}. Similar to the sputtering process, oxidation in the cool post-shock gas has the potential to produce isotope fractionations in SiO. The higher zero-point energy of $^{28}$SiO could potentially produce a non-equlibrium  Rayleigh-type fractionation as SiO is oxidized to SiO$_2$ and condenses into grains. However, even in molecular clouds, the collision frequency between SiO and OH$^{\cdot}$ will be low enough that this effect is likely to be of limited significance.  

In all cases, the clustering of the data representing a wide variety of astrophysical environments from the Galactic center to the outer disk makes large differences in mass fractionation effects seem unlikely.  The possibility for a decoupling of the growth of the two secondary Si isotopes remains.  However, none of these factors could have modified the isotope ratios of SiO sufficiently to alter the conclusion that the variations in $\rm [^{29}SiO]/[^{28}SiO]$ ratios and $\rm [^{30}SiO]/[^{28}SiO]$ ratios across the Galaxy are surprisingly small. 

\section{Conclusions}

Our finding that secondary/primary Si isotope ratios have no detectable variation across the Galaxy within about $20\%$ does not comport with expectations from the large variation in secondary/primary O isotope ratios of $> \sim 900\%$.  Even when accounting for the prediction that the growth of secondary/primary ratios for Si isotopes should be approximately 1/3 that for O over the same range in metallicity, the observed variation is surprisingly small.  The higher $[\rm ^{29}Si]/[^{28}Si]$ and $[\rm ^{30}S]/[^{28}Si]$ ratios of the ISM relative to solar values suggests growth of Galactic secondary isotopes over the last 4.6 Gyrs. The modest increase in secondary/primary Si isotope ratios and the lack of a significant gradient with Galactocentric distance may be qualitatively consistent with previous suggestions that the increase in secondary/primary silicon isotope ratios has slowed with the increased influence of Type Ia supernovae. This result is in apparent conflict with the hypothesis that solar Si is substantially and anomalously enriched in $\rm ^{28}Si$ relative to the ISM at the time of the birth of the solar system \citep[e.g., ][]{Young:2011, Alexander:1999}.  In light of these conclusions, a careful reexamination of the Galactic distribution of oxygen isotopes seems well warranted.  

The spread in Si isotope ratios found among mainstream SiC grains is similar to the spread in values seen in the modern Galaxy,  suggesting that the presolar SiC grains may record both temporal \emph{and} spatial evolution of silicon isotope abundances in the presolar Galaxy.  The key to the conundrum of the higher $[\rm ^{29}S]/[^{28}Si]$ and $[\rm ^{30}S]/[^{28}Si]$ ratios of some mainstream SiC grains relative to solar may lie with the spread in grain data  rather than with the solar value.  

\acknowledgments

This work was supported by a grant from the NASA Origins program (NNX10AH35G) and from the NASA Emerging Worlds program (NNX17AE78G).  Dr. Ron Maddalena and the rest of the staff at the GBT are gratefully acknowledged for their continued help and support.  Thoughtful input by an anonymous reviewer of manuscript improved the final product.

%\appendix
\section*{Appendix}

Here we derive Equations (\ref{eq:cd4}) and (\ref{eq:fix4}).  We start with the total power per unit bandwidth $\mathcal{P}_{\nu}$ collected by an antenna with a geometric aperture $A_\mathrm{g}$ and aperture efficiency $\eta_\mathrm{a}$.  $\mathcal{P}_{\nu}$ is given by the convolution of the photon occupation number of the source $n_{\gamma}$, and the normalized power pattern of the telescope $P_\mathrm{n}$
\begin{equation} \label{eq:at1}
\mathcal{P}_{\nu} = \frac{A_{\rm g} \eta_{\rm a} h \nu_{u \ell}^3}{c^2} \int\int n_{\gamma}(\theta, \phi) P_{\rm n} (\theta, \phi) \, \sin{\theta}\,  {d}\theta \,  {d}\phi.
\end{equation}
If it is assumed that the source is an isothermal, radially uniform disk, then the double integral in (\ref{eq:at1}) reduces to $\Omega_{\rm s} n_{\gamma}$, and the main beam temperature of the telescope can be expressed using the Nyquist theorem as
\begin{equation} \label{eq:at2}
T_{\rm mb} = \frac{\mathcal{P}_{\nu}}{k \eta_{\rm mb}} = \frac{A_{\rm g} \eta_{\rm a} h \nu_{u \ell}^3}{k c^2 \eta_{\rm mb}}  \Omega_{\rm s} n_{\gamma},
\end{equation}
where $\eta_{\rm mb}$ is the main beam efficiency. Using the relation $A_{\rm g} \eta_{\rm a}= \lambda^2 / \Omega_{\rm a}$ , this expression reduces to
\begin{equation} \label{eq:at3}
T_{\rm mb} = \frac{\Omega_{\rm s} }{\Omega_{\rm mb}} \frac{h \nu_{u \ell}}{k} n_{\gamma},
\end{equation}
where $\Omega_{\rm mb} = \Omega_{\rm a} \eta_{\rm mb}$ is the solid angle subtended by the main beam of the antenna.

The photon occupation number $n_{\gamma}$ in (\ref{eq:at3}) can be expressed as a solution to the equation of radiative transfer. If it is assumed that there are no additional emission sources in the optical path, then for a transition $v = 0, J = u \to \ell$ with a well defined excitation temperature $T_{\rm ex}$ and optical depth $\tau_{\nu}$, the solution is
\begin{equation} \label{eq:at4}
n_{\gamma} = [n_{\gamma}(T_{\rm ex}) - n_{\gamma}(T_{\rm crf})] \ (1 - {\rm exp}(-\tau_{\nu})).
\end{equation}
Integrating the absorption coefficient along the optical path gives the optical depth of the line profile as a function of frequency.  When the source is isothermal along the optical path, the integral becomes proportional to the total column density of the excited state $N_u$, and using the Einstein $A$ coefficient, the optical depth can be expressed as
\begin{equation}  \label{eq:at5}
\begin{split}
\tau_{\nu} &= \int \limits_{s_0}^{s} k_{\nu}(s') \ \mathrm{d}s' \\
&= \frac{c^2}{8 \pi \nu_{u \ell}^2} N_{u} A_{u \ell} \left [ \frac{n_{\ell} g_{u}}{n_{u} g_{\ell}} - 1 \right ] \phi(\nu),
\end{split}
\end{equation}
where  $g_u$ and $g_{\ell}$ are the degeneracies for the upper and lower states, and $n_u$ and $n_{\ell}$ are the fractional level populations for the upper and lower states.

At this point, it is common to apply the Rayleigh-Jeans approximation. However, for a subthermal population of emitters, $h \nu_{u \ell} / k T_{\rm ex}$ might not be $\ll 1$ and thus the Rayleigh-Jeans approximation may not apply.  Avoiding the Rayleigh-Jeans approximation, the main beam temperature $T_{\rm mb}$ can be written in a form that allows for subthermal excitation explicitly.  We start with the substitution  $n_{\ell} g_{u} / n_{u} g_{\ell} - 1 = \exp (h \nu_{u \ell} / k T_{\rm ex}) - 1 = 1 / n_{\gamma}{(T_{\rm ex})}$ in the expression for optical depth (Equation \ref{eq:at5}).   Inserting Equation (\ref{eq:at4}) into (\ref{eq:at3}) and multiplying by $\tau_{\nu}/\tau_{\nu}$, we obtain
\begin{equation} \label{eq:at6}
\begin{split}
T_{\rm mb} &= \frac{\Omega_{\rm s} }{\Omega_{\rm mb}}  \left(  \frac{h c^2 N_{u} A_{u \ell}}{8 \pi k \nu_{u \ell}} \right) 
 \left [  \frac{n_{\gamma}(T_{\rm ex}) - n_{\gamma}(T_{\rm crf})}{n_{\gamma}(T_{\rm ex})} \right]  \\  
 &\times \left ( \frac{1 - {\rm exp}(-\tau_{\nu})}{\tau_{\nu}} \right) \phi(\nu). 
 \end{split}
\end{equation}
Solving for the total column density yields
\begin{equation} \label{eq:at6a}
\begin{split}
N_{u} & = \frac{\Omega_{\rm mb} }{\Omega_{\rm s} } \left(  \frac{8 \pi k \nu_{u \ell}}{h c^2 A_{u \ell}} \right ) 
 \left [ \frac{n_{\gamma}(T_{\rm ex})}{n_{\gamma}(T_{\rm ex}) - n_{\gamma}(T_{\rm crf})} \right ] \\ 
 &\times T_{\rm mb} \left ( \frac{\tau_{\nu}}{1 - {\rm exp}(-\tau_{\nu})} \right) \phi(\nu)^{-1}. 
 \end{split}
 \end{equation}
The photon occupation numbers $n_{\gamma}(T_{\rm crf})$ and $n_{\gamma}(T_{\rm ex})$ are essentially invariant across the line profile in Equation (\ref{eq:at6a}).  Similarly, the frequency factor can be set equal to the frequency at line center because the frequency variation across the line profile is negligible. Therefore, we can write the total column density in terms of the integral of main beam temperature and optical depth. This equation can be converted to a function of radial velocity ${v_{\rm r}}$ with the relation ${ d}{v_{\rm r}}  /c = { d}\nu / \nu_{u \ell}$, resulting in
\begin{equation} \label{eq:at7}
\begin{split}
&N_{u}  \int_{0}^{\infty } \phi(v_{\rm r}) \, {d}v_{\rm r} = \frac{\Omega_{\rm mb} }{\Omega_{\rm s} } \left(  \frac{8 \pi k \nu_{o,u \ell}^2}{h c^3 A_{u \ell}} \right) \\
& \times \left[ \frac{n_{\gamma}(T_{\rm ex})}{n_{\gamma}(T_{\rm ex}) - n_{\gamma}(T_{\rm crf})} \right]
 \int_{0}^{\infty} \left( T_{\rm mb} \frac{\tau_{v_{\rm r}}}{1 - {\rm exp}(-\tau_{v_{\rm r}})} \right) { d}v_{\rm r} . 
\end{split}
\end{equation}
By definition the integral of the line shape function is unity in Equation (\ref{eq:at7}).  

We examine Equation (\ref{eq:at7}) in the limit of optically thin in comparison with the more realistic situation of a finite optical depth in order to extract the optical depth correction factor $\Lambda$.  Where $\tau_{v_{\rm r}} \rightarrow 0$, representing the optically thin limit,  ${\tau_{v_{\rm r}}}/({1 - {\rm exp}(-\tau_{v_{\rm r} }  } ))\rightarrow 1$ and Equation (\ref{eq:at7}) reduces to
\begin{equation} \label{eq:at8}
N_{u} ^{\rm Thin}= C  \left[ \frac{n_{\gamma}(T_{\rm ex})}{n_{\gamma}(T_{\rm ex}) - n_{\gamma}(T_{\rm crf})} \right] W, 
\end{equation}
where $C=(\Omega_{\rm mb}/\Omega_{\rm s}) (8 \pi k \nu_{o,u \ell}^2/(h c^3 A_{u \ell}))$ and $W$ is the integrated line intensity.  In this case the column density is directly proportional to the integrated line intensity.  For the more realistic case of $\tau_{v_{\rm r}} > 0$ Equation (\ref{eq:at7}) becomes
\begin{equation} \label{eq:at9}
N_{u} = (N_{u}^{\rm Thin}/W)  \int_{0}^{\infty} \left( T_{\rm mb} \frac{\tau_{v_{\rm r}}}{1 - {\rm exp}(-\tau_{v_{\rm r}})} \right) { d}v_{\rm r}.
\end{equation}

Considering that $N_u^{\rm Thin}$ is the ideal case and that $N_u$ is the more general case, their ratio defines the correction factor for optical depth $\Lambda$:
\begin{equation} \label{eq:at10}
\frac{N_u}{N_u^{\rm Thin}} =  \frac{ \displaystyle \int_{0}^{\infty}  T_{\rm mb} \quad \tau_{v_{\rm r}}/\bigl(1 - {\rm exp}(-\tau_{v_{\rm r}})\bigr) dv_{\rm r} }{W} \equiv \Lambda.
\end{equation}
We note that the definition of $\Lambda$ in Equation (\ref{eq:at10}) is equivalent to the ratio $N_u/N_u^{\rm Thin}$ given by \cite{Mangum2015}.  This can be seen by recalling that $T_{\rm mb}$ is a function of $T_{\rm ex}$ that has the form  $f(T_{\rm ex}) (1-\exp({-\tau_{v_{\rm r}})})$ (e.g., Equations \ref{eq:at3} and \ref{eq:at4}).  Mindful of the definition of $W$, substitution into Equation (\ref{eq:at10}) yields
\begin{equation} \label{eq:at11}
\begin{split}
\frac{N_u}{N_u^{\rm Thin}} & = \frac{ \int_{0}^{\infty}  \tau_{v_{\rm r}} dv_{\rm r}}  {\int_{0}^{\infty}(1 - {\rm exp}(-\tau_{v_{\rm r}})\bigr) dv_{\rm r} }\\
&= \Lambda,
\end{split}
\end{equation}
which is Equation (\ref{eq:fix4}).

Comparing Equations (\ref{eq:at8}), (\ref{eq:at9}), and (\ref{eq:at11}) allows us to write the general equation relating column densities to integrated line intensities: 
\begin{equation} \label{eq:at12}
N_{u} = W \Lambda C  \left[ \frac{n_{\gamma}(T_{\rm ex})}{n_{\gamma}(T_{\rm ex}) - n_{\gamma}(T_{\rm crf})} \right]. 
\end{equation}

We are interested in the ratio of isotopologue column densities. The ratios of aperture and main beam efficiencies for the two isotopologues are both very nearly unity and are safely ignored when the difference between the transition frequencies of isotopologues $p$ and $s$ is small. The antenna theorem shows that to a very good approximation the main beam solid angles scale with the inverse of the square of frequency. For two isotopologues $p$ and $s$ we have
\begin{equation}\label{eq:at13}
\frac{\Omega_{\rm mb}^{s} }{\Omega_{\rm mb}^{p}} \approx \left( \frac{\nu_{u \ell}^{p}}{\nu_{u \ell}^{s}} \right)^{2}
\end{equation}
and therefore the ratio of constants $C$ for the two isotopologues is reduced to
\begin{equation}\label{eq:at14}
\frac{C_s}{C_p} = \frac{A_{u \ell}^{p}}{A_{u \ell}^{s}}.
\end{equation}
With Equations (\ref{eq:at14}) and (\ref{eq:at12}), the ratio of column densities for isotopologues $p$ and $s$ becomes
\begin{equation} \label{eq:at15}
\frac{N_{u}^{s}}{N_{u}^{p}} = \frac{W_{s} \Lambda_{s} }{W_{p} \Lambda_{p}} \left( \frac{A_{u \ell}^{p}}{A_{u \ell}^{s}} \right) 
\left[ \frac{1 - n_{\gamma}(T_{\rm crf}) / n_{\gamma}(T_{\rm ex})_{p}} {1 - n_{\gamma}(T_{\rm crf}) / n_{\gamma}(T_{\rm ex})_{s}} \right].
\end{equation}
This equation can be reduced further by expanding the Einstein $A$ coefficient as
\begin{equation}\label{eq:at16}
A_{ul} = \frac{64 \pi^4 \nu_{o,ul}^3}{3 h c^3 g_u} \left | \langle \psi_u | \mathbf{R} | \psi_l \rangle \right | ^2 ,
\end{equation}
where $\left | \langle \psi_u | \mathbf{R} | \psi_l \rangle \right | ^2$ is the transition dipole moment matrix element from state vector $\psi_u$ to state vector $\psi_l$. With this final substitution, Equation (\ref{eq:at15}) becomes
\begin{equation} \label{eq:at17}
\frac{N_{u}^{s}}{N_{u}^{p}} = \frac{W_{s} \Lambda_{s}} {W_{p} \Lambda_{p}} \left( \frac{\nu_{u \ell}^{p}}{\nu_{u \ell}^{s}} \right)^{3}
\left[ \frac{1 - n_{\gamma}(T_{\rm crf}) / n_{\gamma}(T_{\rm ex})_{p}} {1 - n_{\gamma}(T_{\rm crf}) / n_{\gamma}(T_{\rm ex})_{s}} \right],
\end{equation}
\\
which is Equation (\ref{eq:cd4}) in the main text that is used to extract silicon isotopologue ratios.

%\bibliography{Master_copy}
\bibliographystyle{apj}

\end{document}